\documentclass[aps, twocolumn,amsmath,amssymb,floatfix,superscriptaddress]{revtex4-1}
\usepackage[table]{xcolor}
\usepackage{graphicx}
\usepackage{amsmath,fullpage,amssymb}
\usepackage{graphics,epsfig}
\usepackage{makeidx}

\usepackage{titlesec} 
\titleformat{\subsection}[runin]
{\normalfont\bfseries}{\thesubsection{.}}{1em}{}[.]

\titleformat{\section}
{\normalfont\sffamily\large\bfseries}{\thesection{.}}{1em}{\MakeUppercase}

\usepackage[table]{xcolor}
\usepackage{graphicx}
\usepackage{amsmath,fullpage,amssymb}
\usepackage{graphics,epsfig}
\usepackage{makeidx}
\usepackage{gensymb}
\usepackage{siunitx}

\newcommand*{\citen}[1]{
  \begingroup
    \romannumeral-`\x 
    \setcitestyle{numbers}%
    \cite{#1}%
  \endgroup   
}

\def\ln{{\operatorname{ln}}}

\def\rmd{{\mathrm{d}}}

\def\rme{{\mathrm{e}}}

\def\Equation{Equation}
\def\Eq{eq}
\def\Eqs{eqs}
\def\Figure{Figure}
\def\Fig{Figure}

\def\SItext{{\em Supporting Information}}
\def\ie{{\em i.e.}}
\def\eg{{\em e.g.}}

\newcommand{\Av}[1]{{\bf #1}}

\newcommand{\kB}{k_\textrm{B}}
\newcommand{\chgA}[1]{\textcolor{black} {#1}}

\newcommand{\trm}[1]{{\textrm{#1}}}

\usepackage{soul}	

\begin{document}

\title{Selective molecular transport in thermo-responsive polymer membranes: role of nanoscale hydration and fluctuations}


\author{Matej Kandu\v{c}}
\email{matej.kanduc@helmholtz-berlin.de}
\affiliation{\rm\small Research Group for Simulations of Energy Materials, Helmholtz-Zentrum Berlin f\"ur Materialien und Energie, Hahn-Meitner-Platz 1, D-14109 Berlin, Germany}

\author{Won Kyu Kim}
\affiliation{\rm\small Research Group for Simulations of Energy Materials, Helmholtz-Zentrum Berlin f\"ur Materialien und Energie, Hahn-Meitner-Platz 1, D-14109 Berlin, Germany}

\author{Rafael Roa}
\affiliation{\rm\small Departamento de F\'{i}sica Aplicada I, Facultad de Ciencias, Universidad de M\'{a}laga, Campus de Teatinos s/n, E-29071 M\'{a}laga, Spain}
\affiliation{\rm\small Research Group for Simulations of Energy Materials, Helmholtz-Zentrum Berlin f\"ur Materialien und Energie, Hahn-Meitner-Platz 1, D-14109 Berlin, Germany}

\author{Joachim Dzubiella}
\email{joachim.dzubiella@helmholtz-berlin.de}
\affiliation{\rm\small Physikalisches Institut, Albert-Ludwigs-Universita\"at Freiburg, Hermann-Herder Str.\ 3, D-79104 Freiburg, Germany}
\affiliation{\rm\small Research Group for Simulations of Energy Materials, Helmholtz-Zentrum Berlin f\"ur Materialien und Energie, Hahn-Meitner-Platz 1, D-14109 Berlin, Germany}


\pagenumbering{arabic}
\noindent

\parindent=0cm
\setlength\arraycolsep{2pt}

\begin{abstract}
For a wide range of modern soft functional materials the selective transport of sub-nanometer-sized molecules (`penetrants') through a stimuli-responsive polymeric membrane is key to the desired function. In this study, we investigate the diffusion properties of penetrants ranging from non-polar to polar molecules and ions in a matrix of collapsed Poly(N-isopropylacrylamide) (PNIPAM) polymers in water by means of extensive molecular dynamics simulations. We find that the water distributes heterogeneously in fractal-like cluster structures embedded in the nanometer-sized voids of the polymer matrix.  The nano-clustered water acts as an important player in the penetrant diffusion, which proceeds via a hopping mechanism through `wet' transition states: the penetrants hop from one void to another via transient water channels opened by rare but decisive polymer fluctuations.  The diffusivities of the studied penetrants extend over almost five orders of magnitude and thus enable a formulation of an analytical scaling relation with a clear non-Stokesian, exponential dependence of the diffusion coefficient on the penetrant's radius for the uncharged penetrants. Charged penetrants (ions) behave differently as they get captured in large isolated water clusters. Finally, we find large energetic activation barriers for hopping, which significantly depend on the hydration state and thereby challenge available transport theories.  
\vspace{5ex}
\end{abstract}

\maketitle
\setlength\arraycolsep{2pt}

Thermo-responsive polymer systems, typically realized as polymeric liquids, networks, or hydrogels in solutions, have become integral components in the engineering of soft functional materials. Their attractiveness stems primarily from their responsive behavior, high water content, and rubbery nature, being similar to biological tissue~\cite{peppas1997hydrogels}. These versatile and useful features triggered many applications in material science, spanning from drug delivery~\cite{peppas1997hydrogels, stuartNatMat2010, kabanovAngew2009, oh2008development,yingApplicationsSM2011}, catalysis~\cite{jandtCatal2010, ballauffRoyal2009, dzubiellaAngew2012}, biosensing~\cite{yingApplicationsSM2011,ravaineGlucose2006}, thin-film techniques~\cite{yingApplicationsSM2011}, as well as in environmental science~\cite{serpeACS2011}, including nanofiltration\cite{shannon:nature}, water purification, and desalination applications~\cite{Nghiem:environment, desalination}. One of the most commonly studied responsive polymer is poly(N-isopropylacrylamide) (PNIPAM), which exhibits its volume transition in water close to the human body temperature~\cite{pelton2000, hudsonProgPolySci2004, halperin2015poly}. As a versatile model component, it has been prototypical for many developments of soft responsive materials~\cite{stuartNatMat2010, ballauffAngew2006, ballauffSmartPolymer2007}.

In most of the above mentioned applications, the selective diffusive transport of small solutes (`penetrants') through the polymer matrix is key to the desired function. For instance, in nanocarrier particles, the permeability of the polymer shell plays a crucial role in uptake and release rates of functional penetrants~\cite{stuartNatMat2010}. In particular, hydrogels can be used to release small ligands and drugs in a controlled way over time, or as a response to a local chemical or physical stimulus~\cite{hoffman1987applications, gehrke1997factors}. In desalination, selective salt flux and permeability is key to be optimized, which strongly couples to water permeability of the polymer membrane~\cite{desalination}.  Other examples include responsive `nanoreactors'~\cite{ballauffAngew2006, dzubiellaAngew2012, lis-marzan2008, stamm2014}, where a responsive hydrogel shell around the catalysts is used to tune the reactor's selective permeability of reactants and by that the reaction rate~\cite{ballauffChemSocRev2012, roa2017catalyzed}. 

Due to its fundamental importance, the study of the penetrant transport through polymer meshes and gels has a long history. While the diffusion in swollen (dilute or semi-dilute)  polymer regimes has been tackled successfully by a large body of different theories~\cite{masaro1999physical, amsden1998solute}, a much more challenging problem represents the diffusion in collapsed, highly concentrated polymers, such as melts, glasses, and gels. Indeed, due to material-specific complexity on nanoscales, such as specific solute--polymer interactions in aqueous solution~\cite{walter-PNIPAM2010, vegtJPCB2011, stevens_Macro2012, mukherji2013coil,  chiessi2016influence, rodriguez2014direct, micciulla2016concentration,lesch2016properties, adroher2017conformation, kanduc2017selective}, the full understanding of the problem demands insights by rigorous atomistic modeling. Pioneering simulations since the early '90s~\cite{takeuchi1990jump, muller1991diffusion} have shown that transport in a restricting dense polymer matrix can intrinsically differ from those in liquids and dilute systems. In fact, the diffusion of low-mass particles follows a {\sl hopping} motion~\cite{muller1991diffusion,muller1992computational,gusev1994dynamics, muller1998molecular, fritz1997molecular, hahn1999new, hofmann2000molecular,vdVegt2000temperature, neyertz2010carbon}, where the penetrant is constrained within a microscopic cavity for most of the time and only occasionally hops into a neighboring cavity through a transient void that opens between the chains~\cite{sok1992molecular, muller1993cooperative, hofmann2000molecular}. Despite the aforementioned insights, the long-time diffusivity in dense, atom-resolved polymer models is often so low that its characterization poses a serious challenge for atomistic simulations even nowadays~\cite{xi2013hopping}.  Hence, most useful insight on scalings (\eg, the diffusion versus temperature or solute size) have become just recently available from still quite coarse-grained simulations without solvent~\cite{zhang2017molecular, zhang2018coarse} and generic statistical mechanics theories on activated hopping~\cite{cai2015hopping, zhang2017correlated}.

Hence, molecular insights into penetrant diffusion in concentrated polymers including chemical specificity and in particular the influence of hydration has been very scarce. What is of particular concern in responsive polymer solutions or hydrogels is that the hydration state and polymer packing fraction are temperature dependent. As such, temperature interpolates between very wet swollen and much less hydrated collapsed states of the polymer, and it is unclear what role the water plays in the particular regimes. Is there sufficient water such that diffusion is simply Stokes-like as in a homogeneous fluid around obstructions~\cite{masaro1999physical, amsden1998solute}? Or is the diffusion dominated by hopping and the structural rearrangements of polymer segments as in the statistical mechanics pictures of very dense and `dry' polymer melts~\cite{cai2015hopping, zhang2017correlated}? If so, does water contribute to the temperature dependence and activation (free) energies of diffusion, and how?

Experiments indicate that the distribution of water in dense polymers may be highly non-trivial and may crucially contribute to transport and thermodynamic properties of penetrant molecules~\cite{muller1998diffusion}. For instance, it has been indicated that, depending on the polarity of the host polymer matrix, water can either be uniformly distributed~\cite{karlsson2004molecular} or tend to cluster~\cite{fukuda1998clustering}. The existence of disconnected water clusters in collapsed PNIPAM gels was also revealed by dielectric dispersion spectroscopy~\cite{sasaki1999dielectric}. 
Apart from these and other fragmented experimental facts, a bigger picture of the interplay between the polymers, water, and diffusing particles is still missing and is difficult to obtain from experiments. 
Namely, most experimental techniques average over the behavior of a large number of molecules, and thus can elucidate only particular facets of the entire problem. On the contrary, detailed and trustworthy atomistic simulations can offer at least qualitative insights of a broader picture, which we pursue in this work.


In this work, we employ for the first time extensive molecular dynamics simulations to examine the diffusion of small and medium-sized penetrant molecules of various kinds in a collapsed thermo-responsive PNIPAM polymer in water above its lower critical solution temperature~(LCST).  We first analyze sorbed water content in the collapsed polymer and its structure. After that we focus on diffusion of selected penetrant molecules in the polymer at different temperatures and water contents. 
The resulting diffusivities span almost over five orders of magnitude and hence enable us to extract reliable scaling laws to be compared to previous work. Here, sorbed nano-clustered water
and matrix fluctuations play an important role in the diffusion by creating pathways for penetrant molecules. We finally analyze the hopping diffusion mechanisms in terms of energy barriers in relation to solute size and hydration state of the polymer.

\section*{Results and discussions}

\subsection*{Water content in the collapsed phase}
\begin{figure*}[t]\begin{center}
\begin{minipage}[b]{0.33\textwidth}\begin{center}
\includegraphics[width=\textwidth]{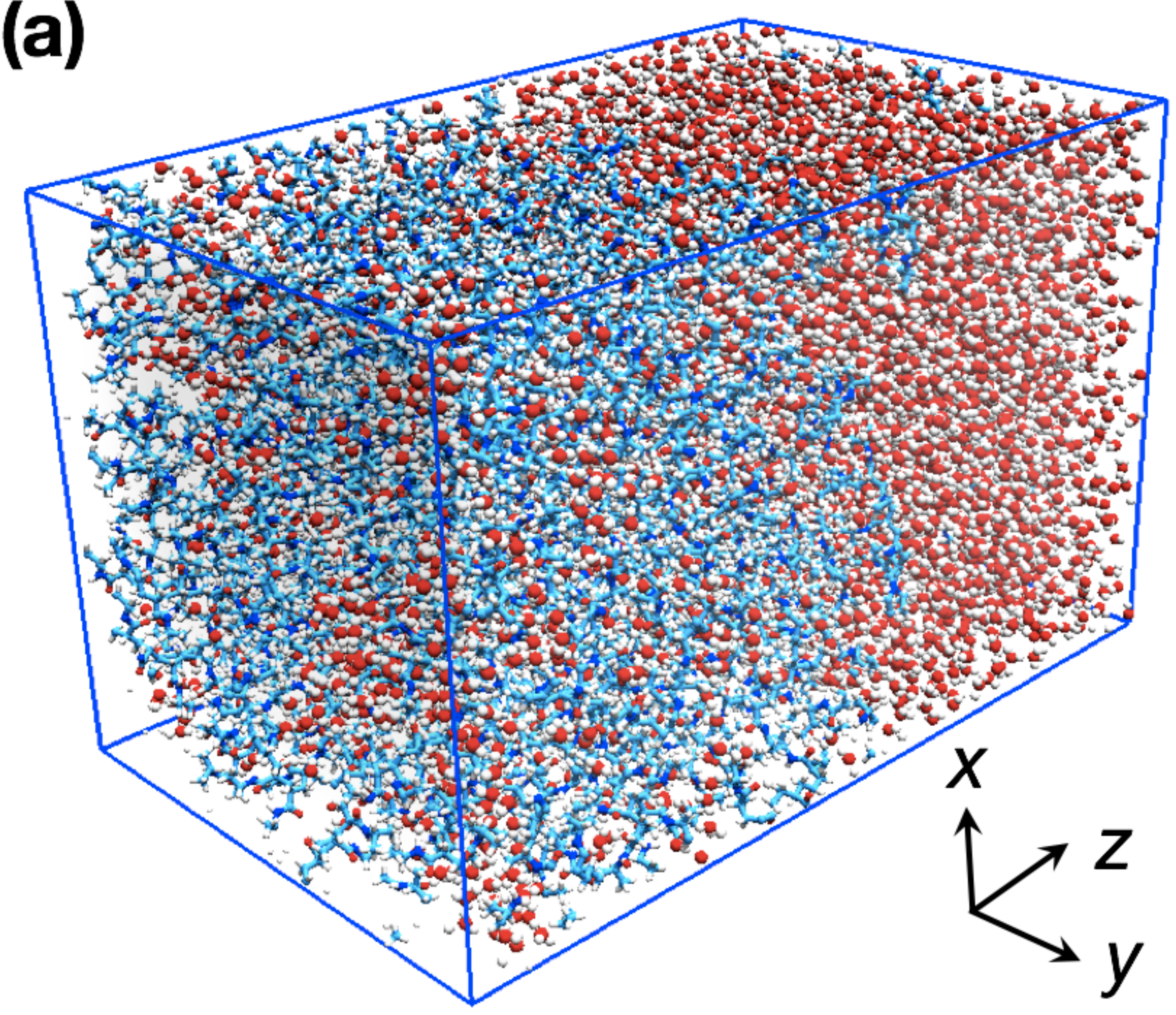}
\end{center}\end{minipage}\hspace{3ex}
\begin{minipage}[b]{0.30\textwidth}\begin{center}
\includegraphics[width=\textwidth]{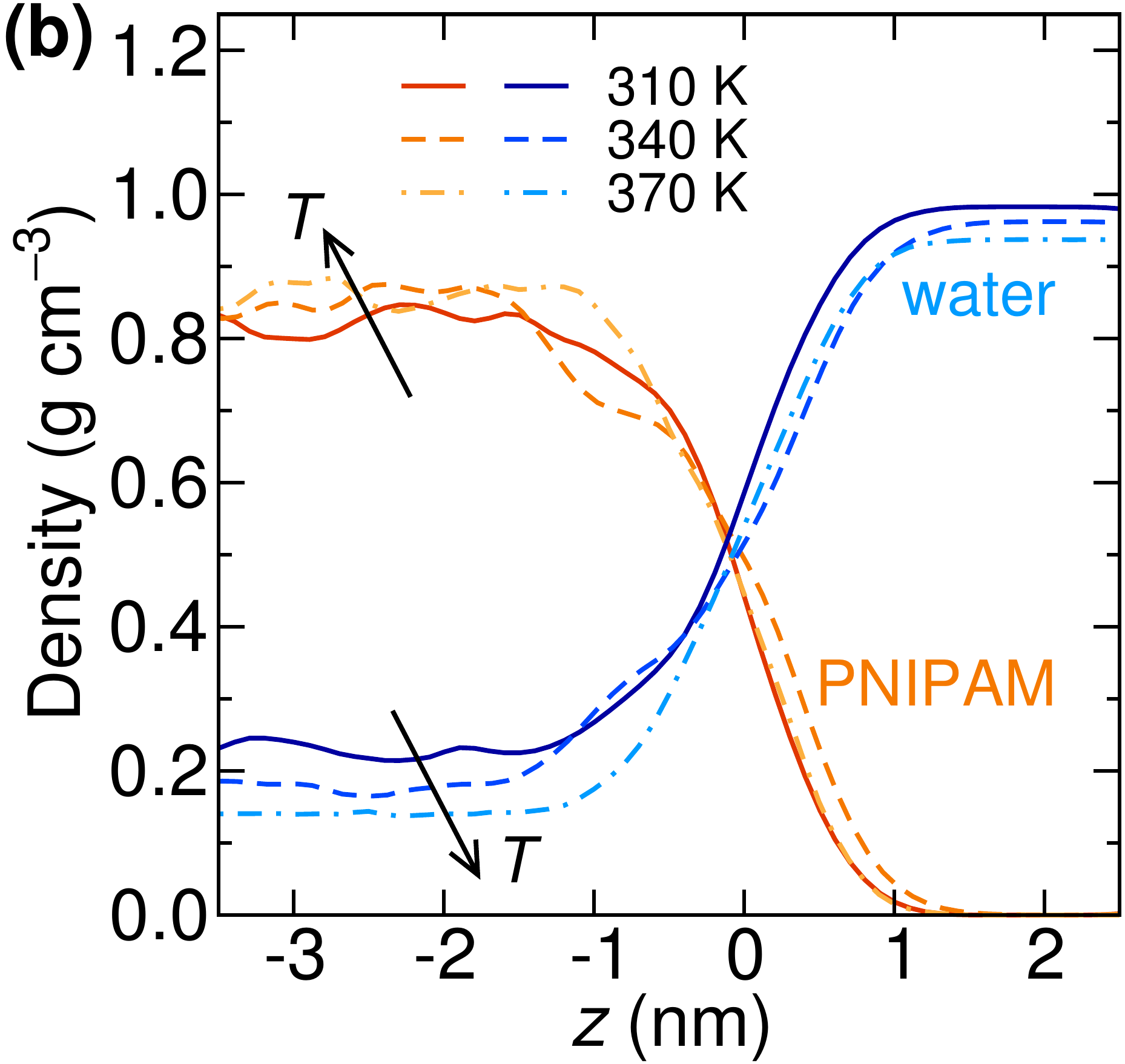}
\end{center}\end{minipage}\hspace{3ex}
\begin{minipage}[b]{0.29\textwidth}\begin{center}
\includegraphics[width=\textwidth]{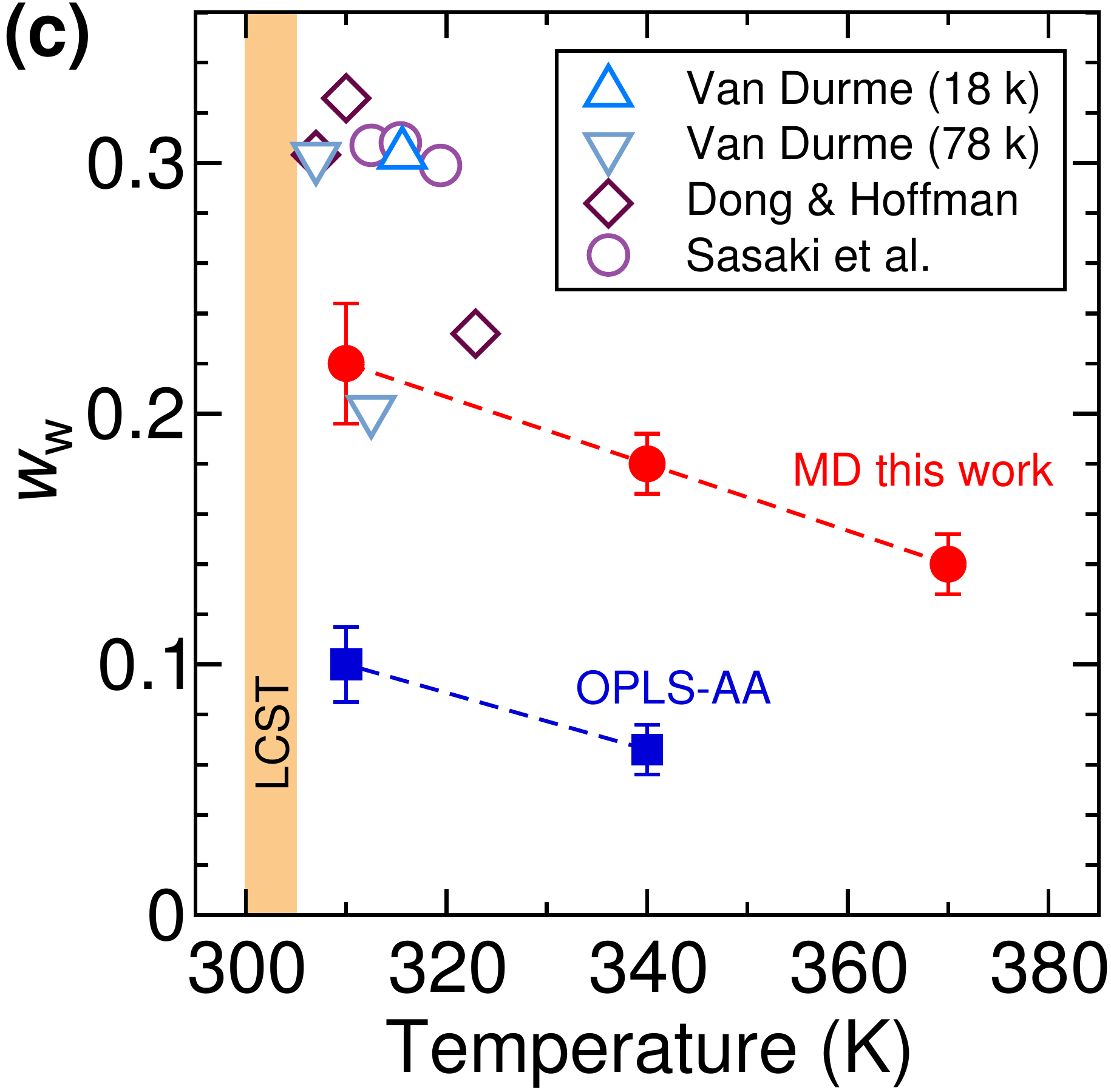}
\end{center}\end{minipage}
\caption{Two-phase system. (a) A snapshot of the collapsed PNIPAM slab, composed of 48 atactic PNIPAM polymers (in blue color), in contact with a bulk water phase. (b)~Mass density profiles of water and PNIPAM across the slabs at different temperatures, where the origin of $z$-axis is set at the Gibbs dividing surface of the water phase. (c)~Mass fraction of water in the polymer phase as a function of temperature (red circles connected by straight dashed lines). Results of the same simulation setup obtained with the OPLS-AA force field~\cite{opls1988} (not used in this study) are shown by blue square symbols. 
Open symbols show various experimental measurements. Triangles are the PNIPAM/water coexistence (binodal) curves by Van Durme et al.~\cite{vanDurme2004kinetics} for molecular weights of 18 and 78 kDa. (The published data correspond to cloud-point temperatures and subject to demixing hysteresis~\cite{halperin2015poly}. For the purposes of comparison, they are shifted by 8~K, such that the lowest demixing value coincides with the LCST). Diamonds and circles represent the water content in PNIPAM hydrogels (\ie, cross-linked networks) by Dong and Hoffman~\cite{dong1990synthesis} and \chgA{Sasaki} et al.~\cite{sasaki1999dielectric}, respectively.
}
\label{fig:rwT}
\end{center}\end{figure*}


To set up a system of collapsed PNIPAM polymers (20-monomer-long atactic chains) above the LCST with the same chemical potential of water in as in bulk water, we first construct a system with two distinct phases~\cite{botan2016direct, adroher2017conformation}: a collapsed amorphous polymer phase in one part of the box, forming a membrane, and a polymer-free water reservoir in the other, as shown in \Fig~\ref{fig:rwT}a. We use a novel, recently introduced, OPLS-based force field~\cite{palivecheyda2018} for the PNIPAM polymers, which better captures the thermo-responsiveness than the standard OPLS-AA~\cite{opls1988}. The simulation details and equilibration procedures are described in the Methods section.

\Figure~\ref{fig:rwT}b shows equilibrated mass density profiles of water (blue shades) and  the polymer (orange shades) at three different temperatures above the LCST. The water density in the polymer phase is drastically reduced compared with the bulk phase. Upon heating, the polymer expels even more water into bulk, which reflects its increasing hydrophobic character with rising temperature. 
In \Fig~\ref{fig:rwT}c, the densities inside the polymer phase have been converted into mass water fraction, $w_\trm w$, and are plotted as a function of temperature (red circles). As seen, the mass fraction roughly linearly decreases with temperature from around $w_\trm w=$~0.22 near the LCST to around 0.14 at 370~K. 
However, the quantitative consensus regarding the experimental PNIPAM/water phase diagram is limited due to various issues facing experimental measurements, such as differences in synthesis protocols, dependence on polymer chain length, and choices of criteria for the onset of demixing~\cite{vanDurme2004kinetics, halperin2015poly}. In \Fig~\ref{fig:rwT}c, for instance,  we also show measurements by Van Durme et al.~\cite{vanDurme2004kinetics} (triangles) for two different molecular weights of PNIPAM (18 and 78 kDa), which differ considerably in the water content, thereby demonstrating the subtlety and sensitivity of the system  on various parameters.
The water content in the collapsed PNIPAM phase can also be compared to cross-linked PNIPAM hydrogels above the transition temperature. The reported values for hydrogels~\cite{dong1990synthesis, sasaki1999dielectric} shown in the figure tend to be a bit higher than the MD values, with $w_\trm w\approx$~0.3 near the LCST. It should be noted that an exact comparison to the hydrogel experiments turns out to be difficult due to possible additional effects of crosslinkers (around 1~wt\% in both experiments), which make the whole network structure less flexible and likely more porous~\cite{raccis2011probing, kaneko1995temperature, dong1986thermally}. However, within these uncertainties, we conclude that the used novel PNIPAM model\cite{palivecheyda2018} captures the overall features of the simulated system reasonably well.
On the same plot, we also show MD results obtained with the standard OPLS-AA~\cite{opls1988} force field of the same system by blue squares. With far lower values of $w_\trm {w}$ than reported by the experiments, we do not find it suitable for our system.
  


\subsection*{Water structure in the collapsed polymer}
\begin{figure}\begin{center}
\begin{minipage}[b]{0.23\textwidth}\begin{center}
\includegraphics[width=\textwidth]{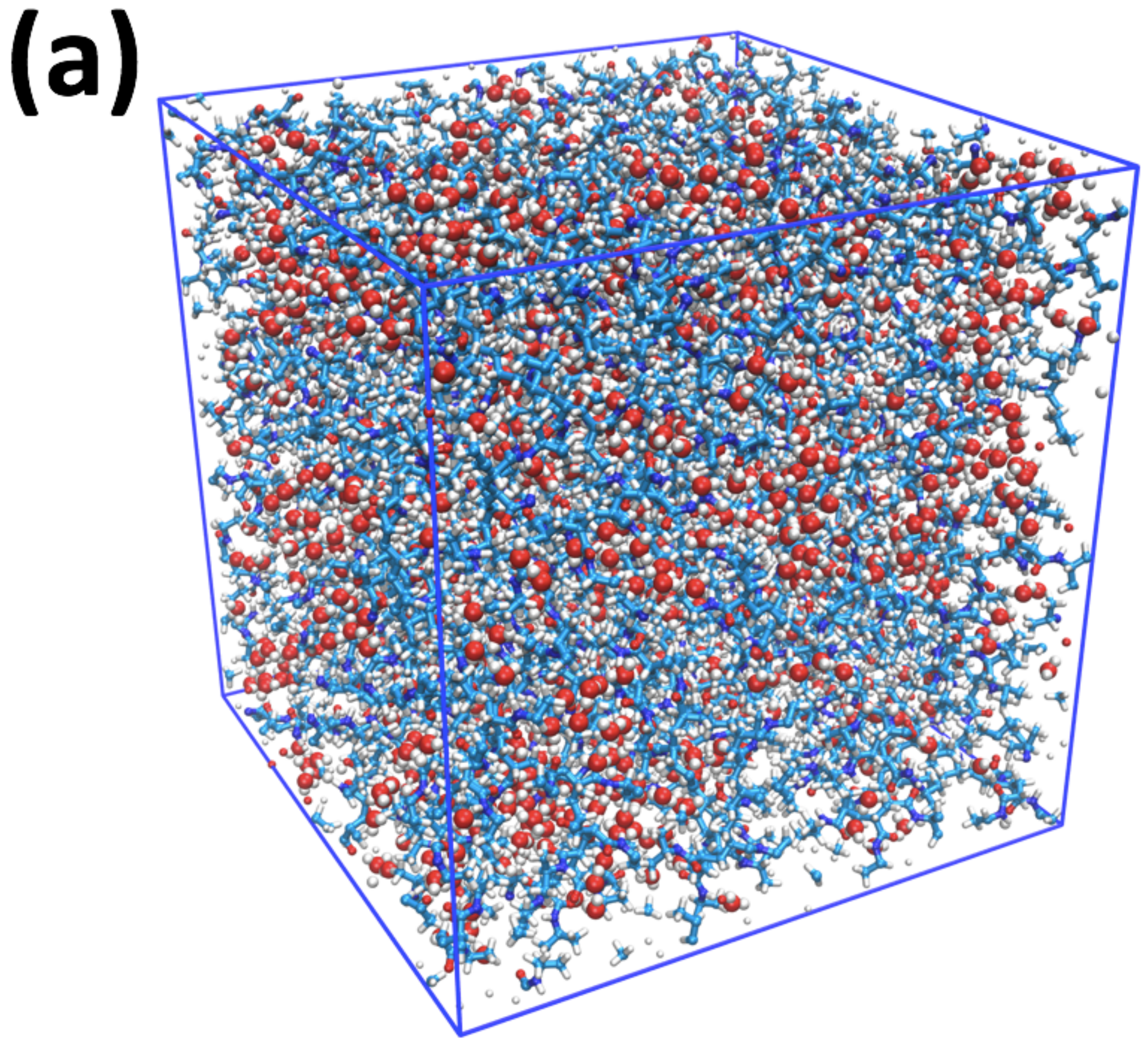} 
\end{center}\end{minipage}
\begin{minipage}[b]{0.23\textwidth}\begin{center}
\includegraphics[width=\textwidth]{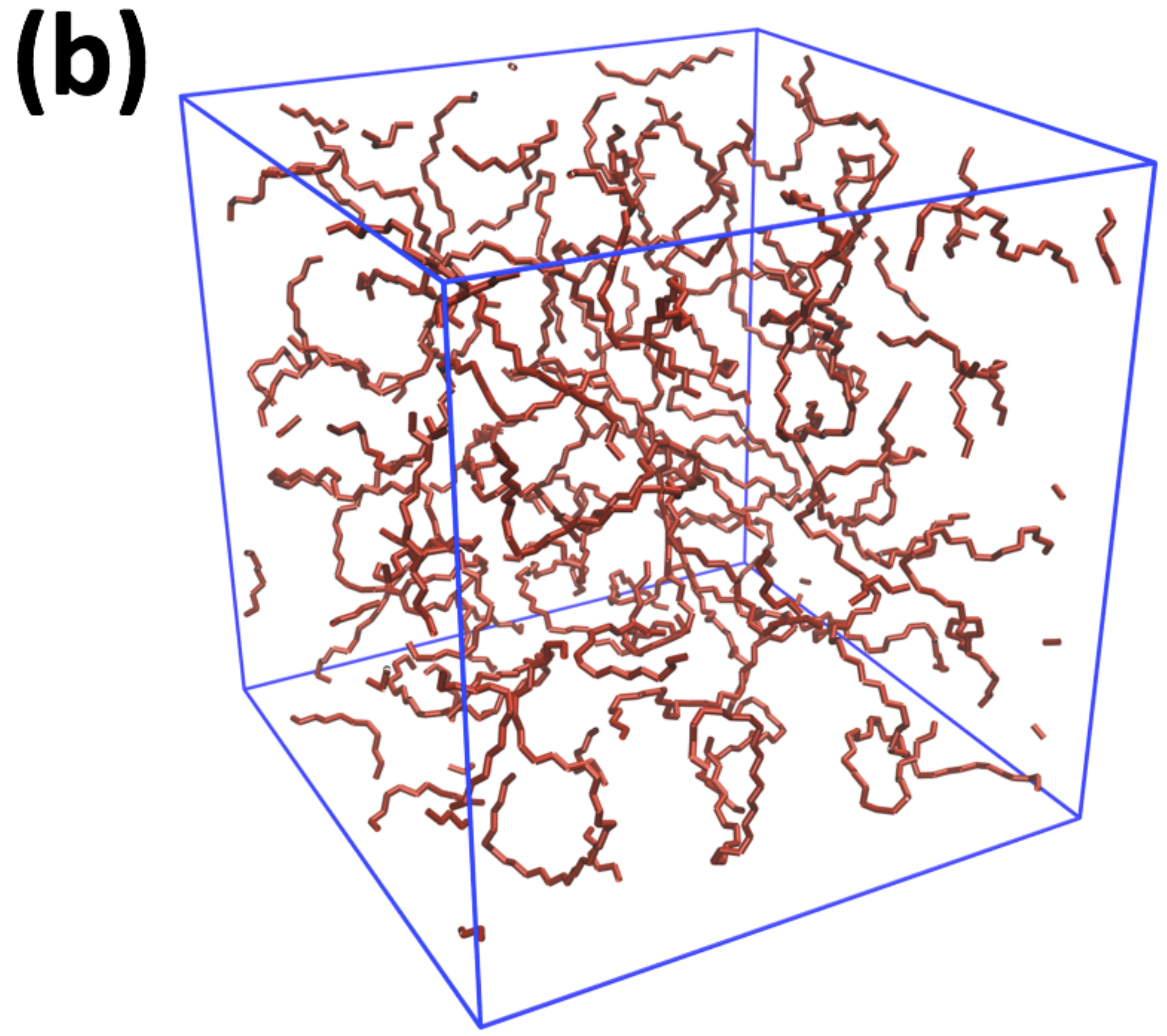} 
\end{center}\end{minipage}
\begin{minipage}[b]{0.23\textwidth}\begin{center}
\includegraphics[width=\textwidth]{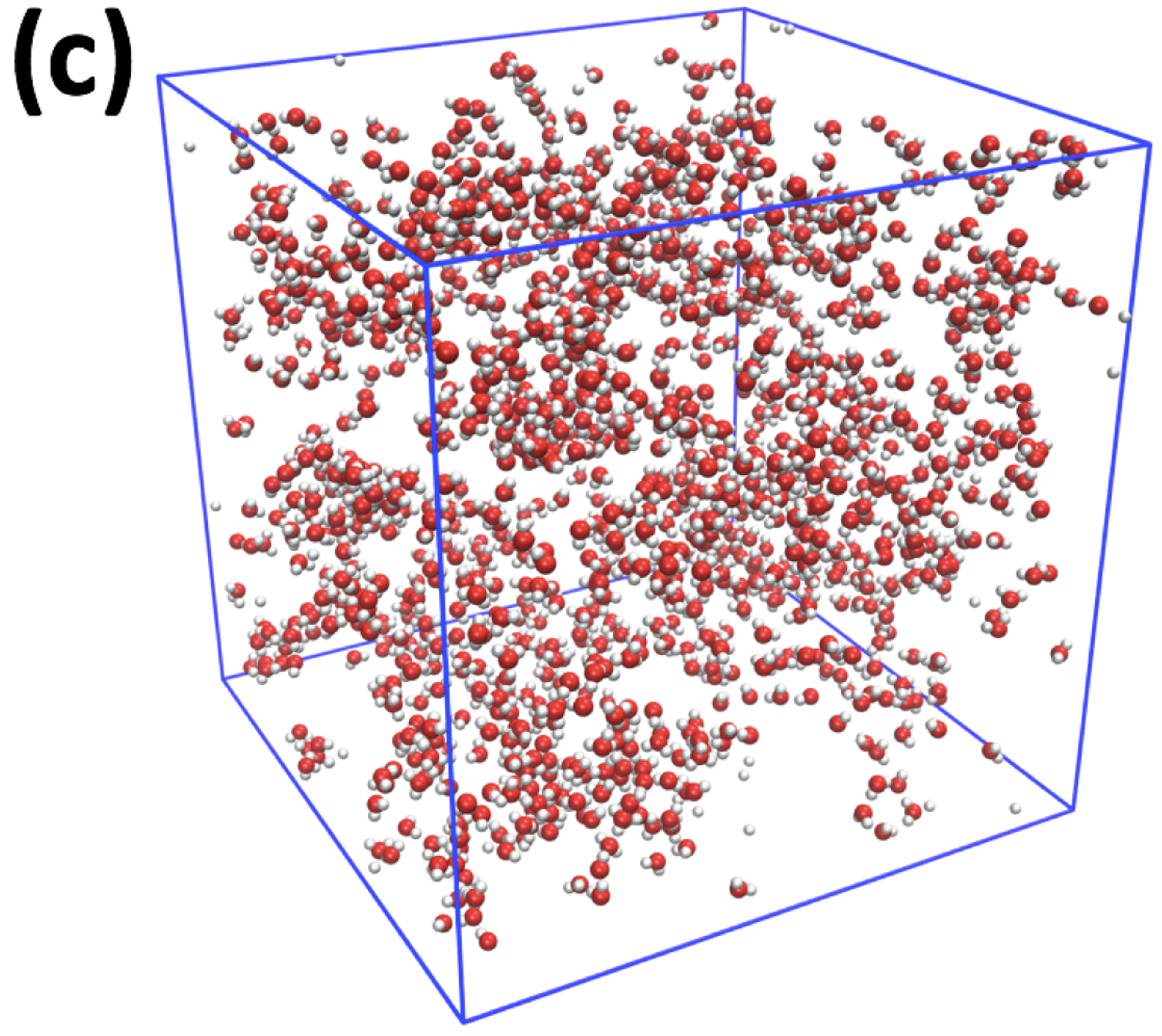} 
\end{center}\end{minipage}
\begin{minipage}[b]{0.23\textwidth}\begin{center}
\includegraphics[width=\textwidth]{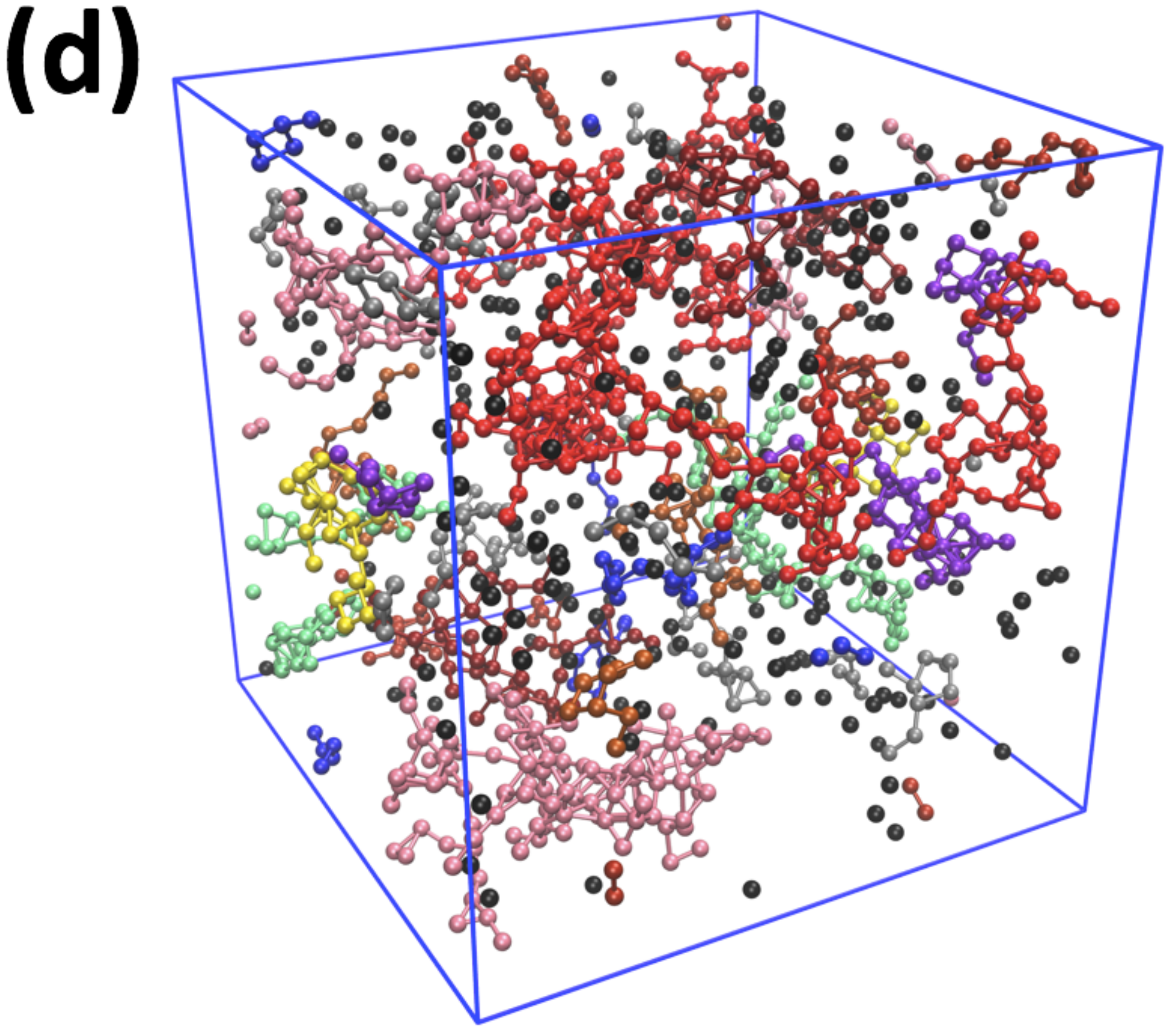} 
\end{center}\end{minipage}
\caption{Snapshot of a single-phase simulation at 340~K, highlighting
(a)~all atoms of the system, (b)~PNIPAM backbones, (c)~water molecules, and (d)~individual water clusters distinguished by different colors (shown as connected water oxygen atoms). The simulation box (blue frame) is approximately cubic with the size $L\approx 6$~nm.}
\label{fig:gelly}
\end{center}\end{figure}

From the above introduced two-phase system (\Fig~\ref{fig:rwT}a) we now switch to a system with a collapsed polymer phase occupying the entire simulation box (\Fig~\ref{fig:gelly}a). Here, the water amount is set in accordance with the outcomes of the two-phase model. Due to periodic boundary conditions, it mimics an infinite `bulk' of collapsed PNIPAM polymers in water or a collapsed weakly cross-linked gel.  The simulation details are described in the Methods section.

A typical simulation snapshot in various representations is shown in \Fig~\ref{fig:gelly}. Panel (a) displays all atoms in the space-filling representation, with PNIPAM chains in blue and water in white--red. Panel (b) highlights the PNIPAM backbones, thereby revealing a disordered amorphous structure with voids between the chains that can accommodate water molecules, displayed in (c) and (d).
As can be noted, water molecules are not uniformly distributed throughout the polymer phase. Instead, they tend to structure into heterogeneous formations, that is, into nanosized water clusters and pockets, as featured in panel (d), where individual water clusters (see further below for the definition of a cluster) are depicted in different colors. 

\begin{figure*}[t]\begin{center}
\begin{minipage}[b]{0.31\textwidth}\begin{center}
\includegraphics[width=\textwidth]{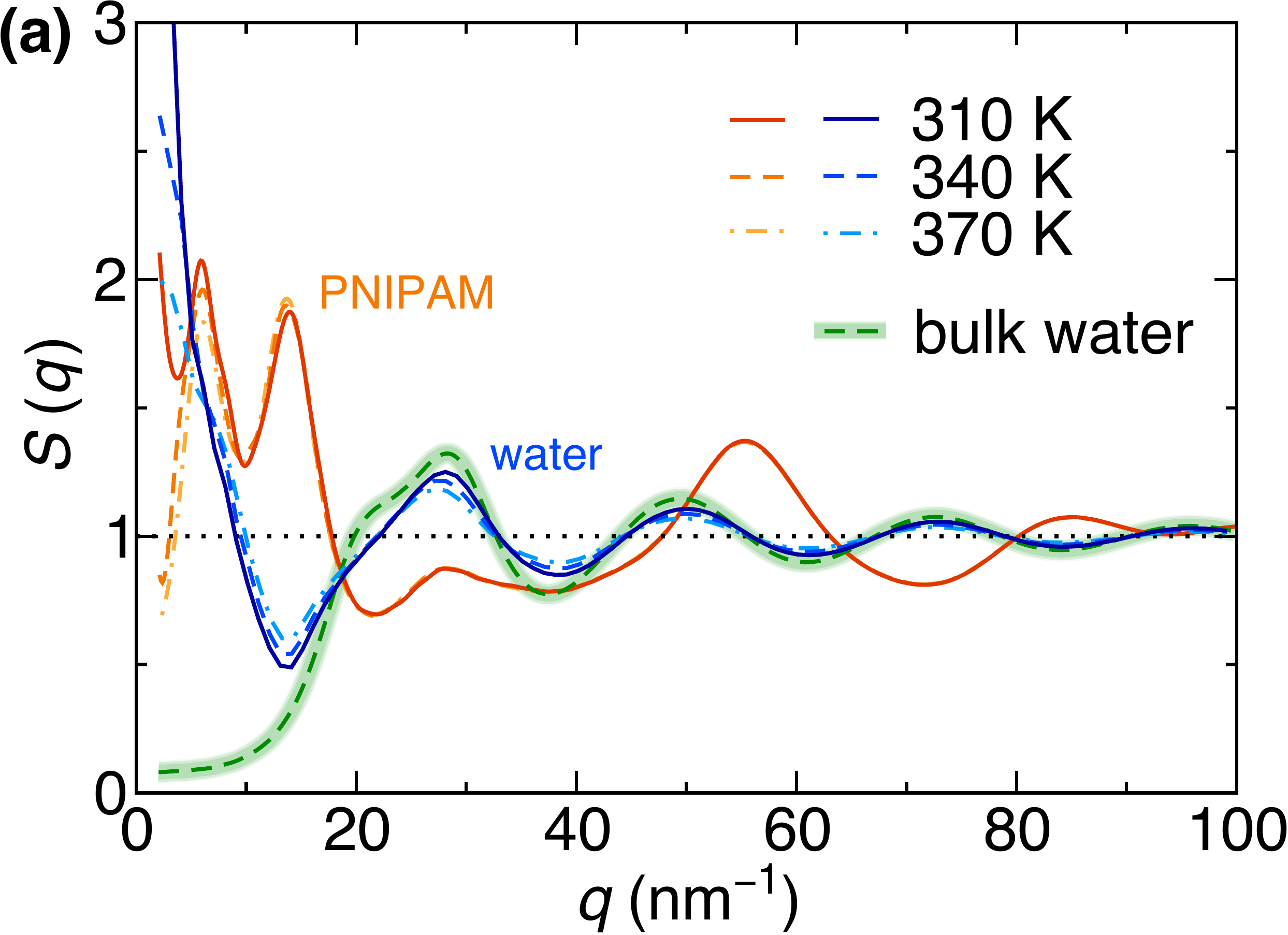}
\end{center}\end{minipage}\hspace{2ex}
\begin{minipage}[b]{0.33\textwidth}\begin{center}
\includegraphics[width=\textwidth]{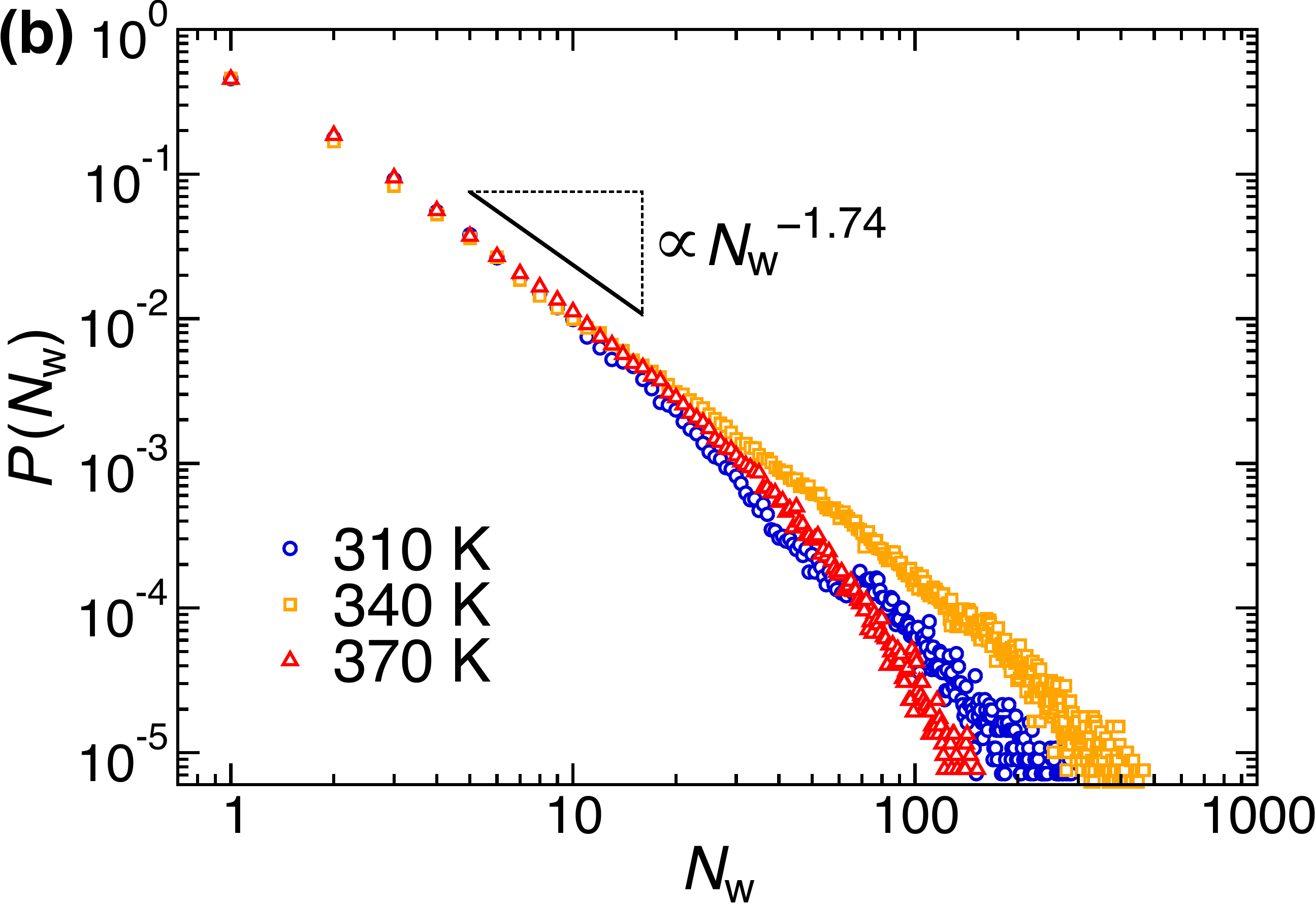}
\end{center}\end{minipage}\hspace{2ex}
\begin{minipage}[b]{0.31\textwidth}\begin{center}
\includegraphics[width=\textwidth]{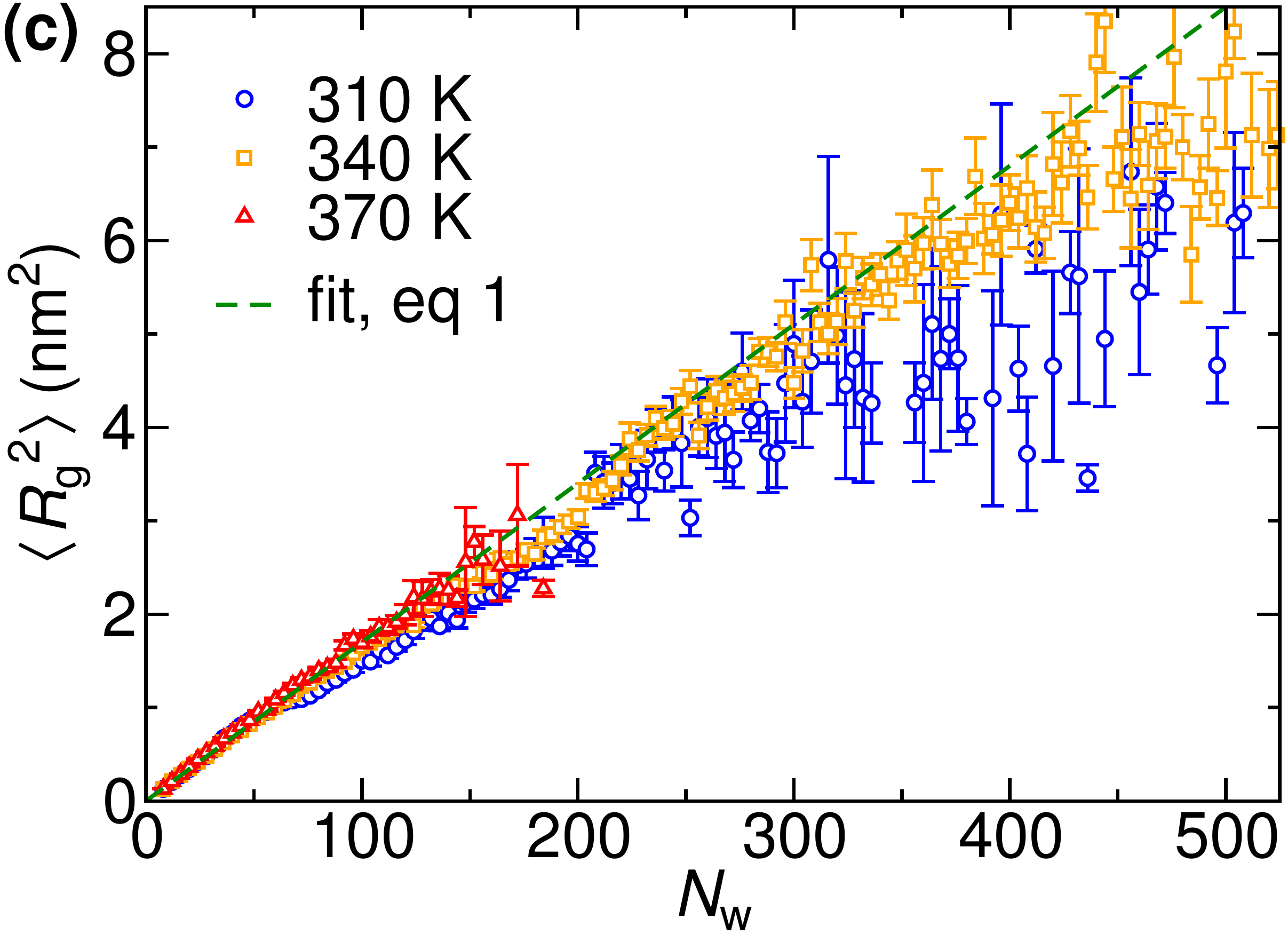}
\end{center}\end{minipage}
\caption{(a)~Structure factor of water (blue shades) and PNIPAM polymers (orange shades) in the collapsed phase at different temperatures. The structure factor of bulk water at 340~K is shown by a green dash-dotted line for comparison.
(b)~Water cluster size distribution in the collapsed polymer at different temperatures.
 (c)~Scaling plot of the mean square radius of gyration $\langle {R_g}^2 \rangle$ of the clusters versus the cluster size in terms of $N_{\rm w}$. A clear linear scaling (cf.\ \Eq~\ref{eq:Rg_scaling}) is revealed (dashed line).}
\label{fig:structure}
\end{center}\end{figure*}

In \Fig~\ref{fig:structure}a we plot the static structure factors of water oxygen atoms (blue shades) and PNIPAM heavy atoms (orange shades) that result from the simulations at different temperatures. For comparison, the structure factor of bulk (SPC/E) water at 340~K is plotted by a green dash-dotted line. The structure factor of water in the polymer differs only slightly from that of bulk water for higher wavenumbers, $q\gtrsim 20$~nm$^{-1}$. Only the disappearance 
of the characteristic shoulder for water at $q\simeq 20$~nm$^{-1}$ indicates some distortion of the tetrahedral hydrogen-bond network~\cite{sorenson, sedlmeier2011spatial}. At lower values, particularly for $q<10$~nm$^{-1}$ (\ie, for length scales $2\pi/q \gtrsim 0.6$~nm), the structure factor in the polymer phase reaches much higher values than in bulk, which must be ascribed to large-scale spatial correlations due to the clustering. Here, also the structure factor of the polymer shows large peaks and reflects the mesh and void structure on the scale of $\simeq$~1~nm, which accommodates the water clusters. Increasing temperature mostly affects the signal at low $q$-values and signifies changes of the nanosized cluster distribution. 

For purposes of further analysis, we define a cluster as a group of the water molecules that are mutually separated by less than  0.35~nm (corresponding to the size of the first water hydration shell~\cite{spce}).
Figure~\ref{fig:structure}b shows the cluster-size distribution $P(N_\trm w)$, defined as the fraction of clusters in terms of their number size $N_\trm w$, in the log--log scale at different temperatures. As can be seen, the clusters do not have a characteristic size but are extremely polydisperse. The size distribution of smaller clusters (with $N_\trm w$ below 20) can be roughly described by the power law $P(N_\trm{w})\propto N_\trm{w}^{-1.74}$.
Larger clusters (with $N_\trm w > 20$) tend to be progressively less present than assumed by the power law.
\chgA{Note a non-monotonicity of the distribution of larger clusters with rising temperature, which could be due to competing effects of varying the water content and increasing hydrophobicity of the polymer matrix.}


A glance at the snapshot in \Fig~\ref{fig:gelly}d reveals that the clusters are far from being compact structures, but rather of `lacy' forms. 
It is known that various aggregation processes (\eg, in colloidal systems) can lead to random cluster formations describable as fractals~\cite{forrest1979long,smirnov1990properties, sorensen1997prefactor}.
A consequence of the fractal morphology is that the cluster size, typically expressed in terms of its radius of gyration $R_\trm g$, scales with the number of particles $N_\trm w$ as $R_\trm g\sim {N_\trm w}^{1/D_\trm f}$, where $D_\trm f$ is fractal dimension~\cite{smirnov1990properties}. 
 Lower values of $D_\trm f$ are associated with more open structures of clusters, and higher values with more compact clusters.
The plot in \Fig~\ref{fig:structure}c, showing the evaluated mean square of $R_\trm g$ versus the cluster size $N_\trm w$, reveals a clear linear relation for not too large clusters,
\begin{equation}
\langle{R_\trm{g}}^2\rangle={\xi_R} N_\trm{w}
\label{eq:Rg_scaling}
\end{equation}
with $\xi_R=0.017$~nm$^2$ (determined by the fit to the MD values for $N_\trm w<75$), shown by a dashed line.
The fractal dimension of the clusters is hence $D_\trm f\simeq 2$, that is, essentially the same as for the random walk.
Larger clusters, however, deviate from the above scaling and tend to be more compact, namely, smaller than predicted by the scaling relation. In the limiting case of completely compact (\ie, spherical) clusters, we expect ${R_\trm g}^2$ to scale as ${N_\trm w}^{2/3}$.

\subsection*{Molecular penetrant hydration}

We examine different types of penetrant molecules, categorized into three major groups: non-polar molecules, polar molecules, and ions (shown in \Fig~\ref{fig:solutes}).
The non-polar molecules comprise noble-gas atoms, simple alkanes, and simple aromatics.
The polar molecules are water, methanol, and 4-nitrophenol~(NP$^0$).
Finally, we treat three monatomic ions and 4-nitrophenolate (NP$^-$), which is the deprotonated form of NP$^0$.
The used force fields are described in the Methods section.

\begin{figure}[h]\begin{center}
\begin{minipage}[b]{0.42\textwidth}\begin{center}
\includegraphics[width=\textwidth]{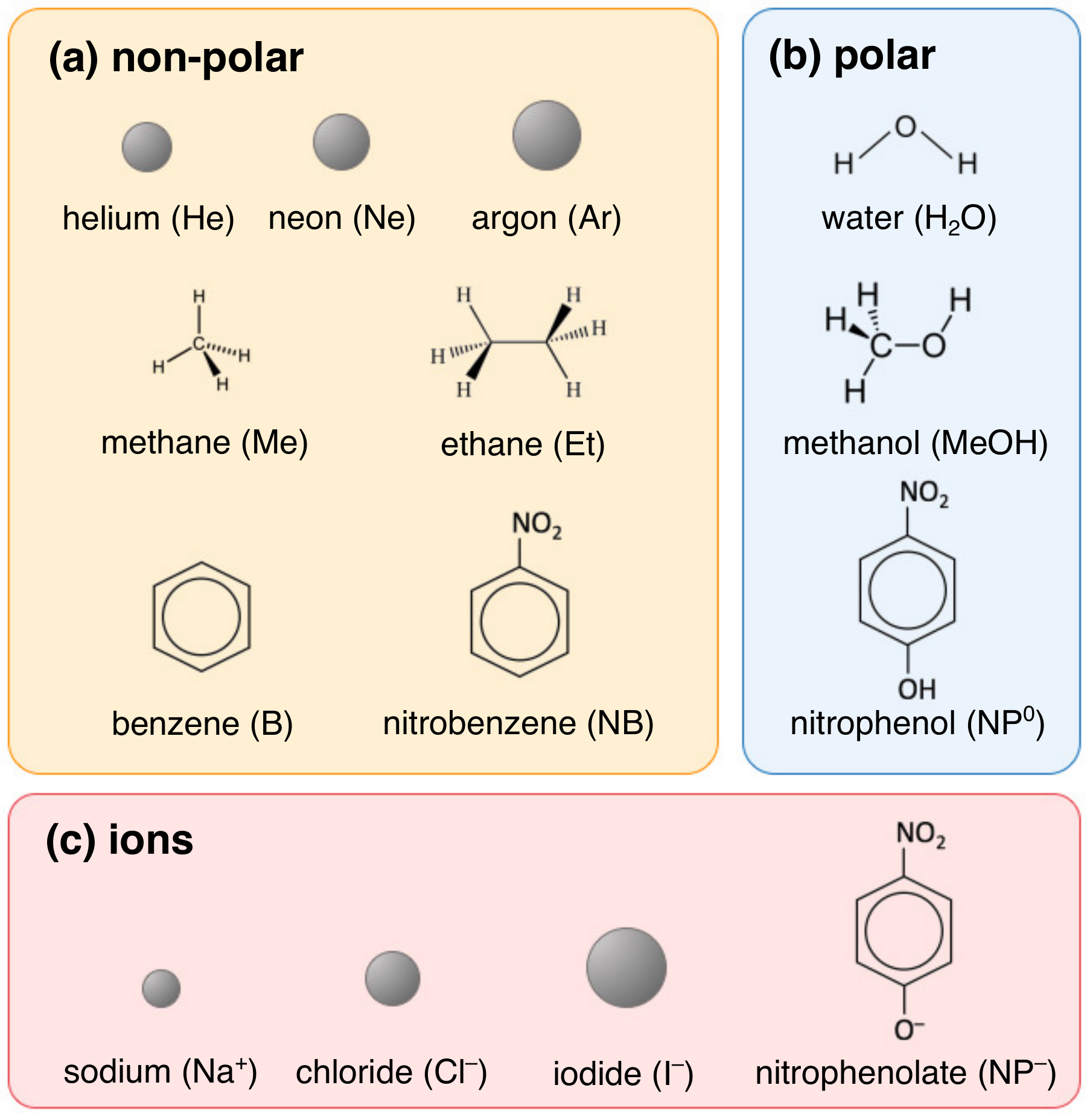}
\label{fig:solutes}
\end{center}\end{minipage}
\caption{The penetrants studied in the simulations: (a)~non-polar molecules, (b)~polar molecules, and (c)~ions.}
\label{fig:solutes}
\end{center}\end{figure}

\begin{figure}[h]\begin{center}
\begin{minipage}[b]{0.43\textwidth}\begin{center}
\includegraphics[width=\textwidth]{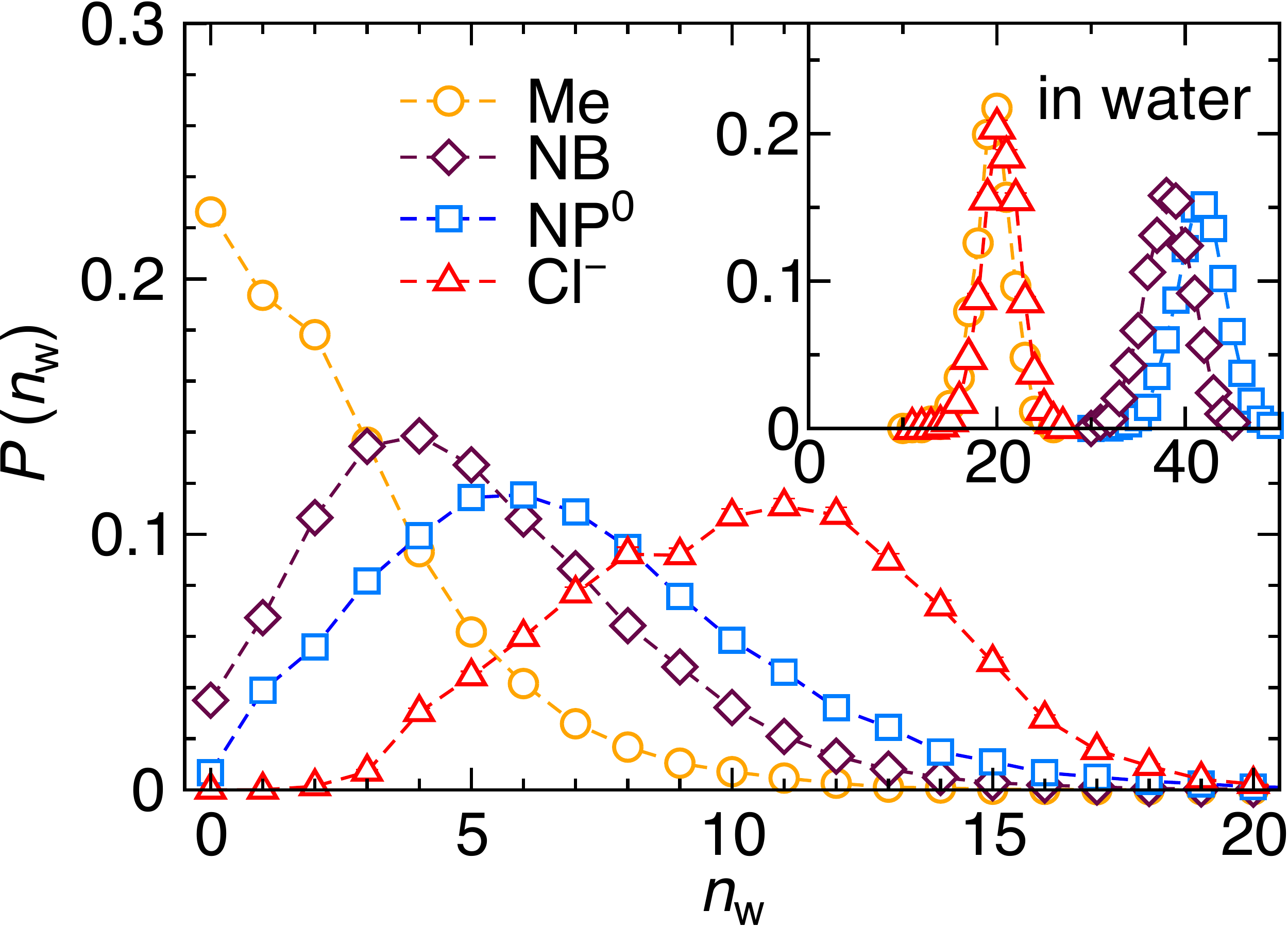}
\end{center}\end{minipage}
\caption{Hydration of penetrants in the polymer phase in terms of probability distribution of number of water molecules $n_\trm{w}$ within 0.54~nm distance around any of the penetrant's atoms. Inset: the probability distributions of the same penetrants in bulk water.}
\label{fig:Pnw}
\end{center}\end{figure}

We first explore the hydration environment in which the penetrants diffuse. For that, we monitor the average hydration number $n_\trm w$, defined as the number of water molecules that reside within a spherical shell of radius $r_\trm{c}=$~0.54~nm (corresponding to the first hydration shell of CH$_4$ in pure water, see \SItext) around any of the penetrant's atoms. 
In \Fig~\ref{fig:Pnw} we show the probability distributions $P(n_\trm{w})$ of the hydration numbers for several penetrants in the collapsed polymer: The non-polar solutes (Me and NB) exhibit a rather weak hydration, with a significant probability of even a completely dry environment ($n_\trm w=0$). 
On the other hand, a polar NP$^0$ is notably more hydrated than the similarly sized NB. 
An extreme case of the hydration behavior is exhibited by ions (shown for Cl$^-$), which are `wrapped' in a considerable hydration layer that they not give up readily. From experimental conductivity measurements, it was already speculated that small ions in a collapsed PNIPAM gel get captured in isolated water clusters~\cite{sasaki1999dielectric}. 
\chgA{The above results also suggest that since ions tend to preferably reside well inside the  water clusters and non-polar molecules in dry regions, the polar molecules may then preferentially reside at the boundaries of the water clusters with their non-polar parts pointing towards the polymer and the polar parts towards water.}
For comparison, we plot in the inset also the hydration numbers of the same molecules in bulk water. The comparison indeed reveals a much larger dehydration of non-polar molecules when transferred from bulk water into the polymer phase.

\subsection*{Molecular penetrant diffusion}

\begin{figure}[h!]\begin{center}
\begin{minipage}[b]{0.34\textwidth}\begin{center}
\includegraphics[width=\textwidth]{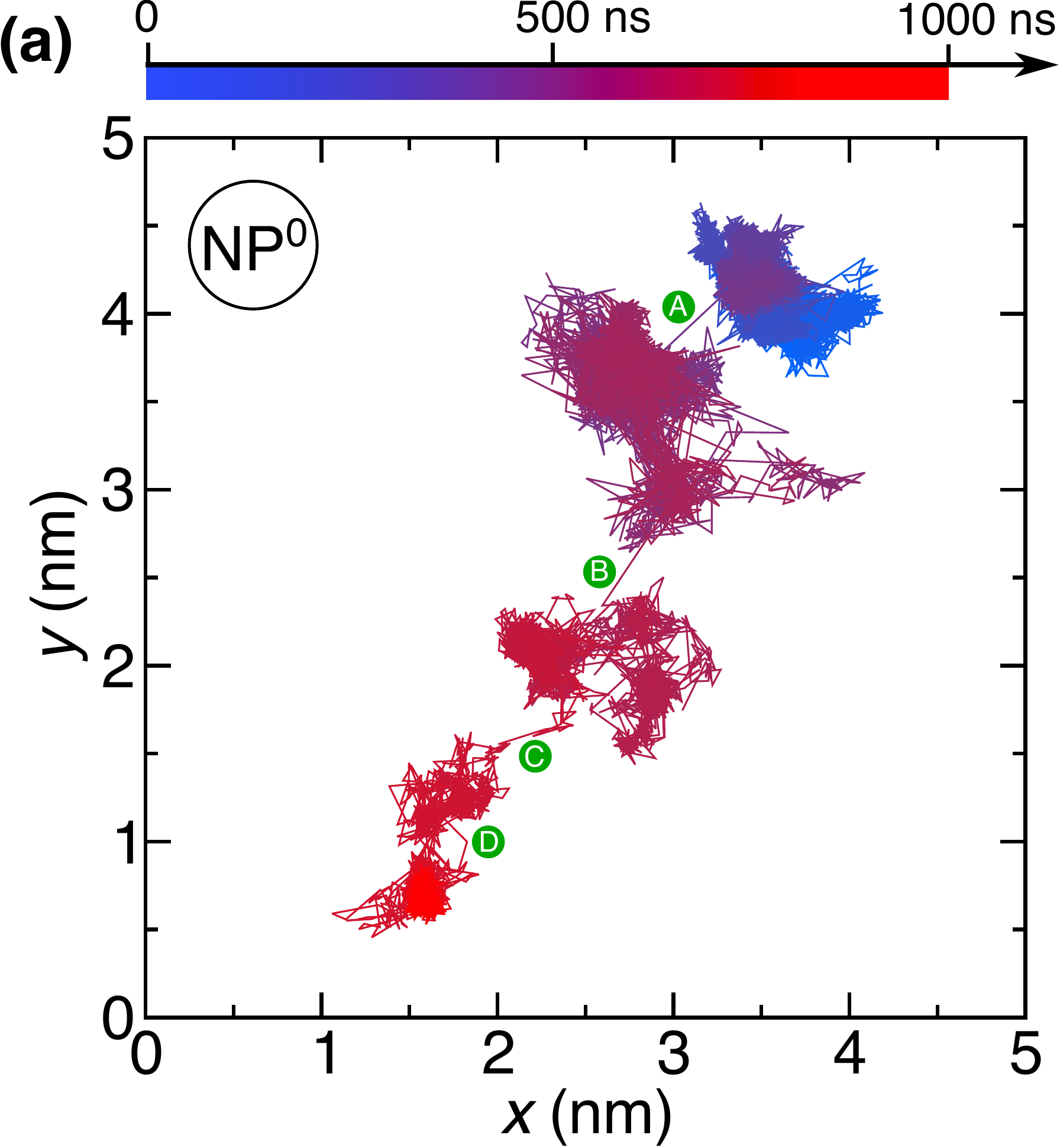}
\end{center}\end{minipage}
\begin{minipage}[b]{0.36\textwidth}\begin{center}
\includegraphics[width=\textwidth]{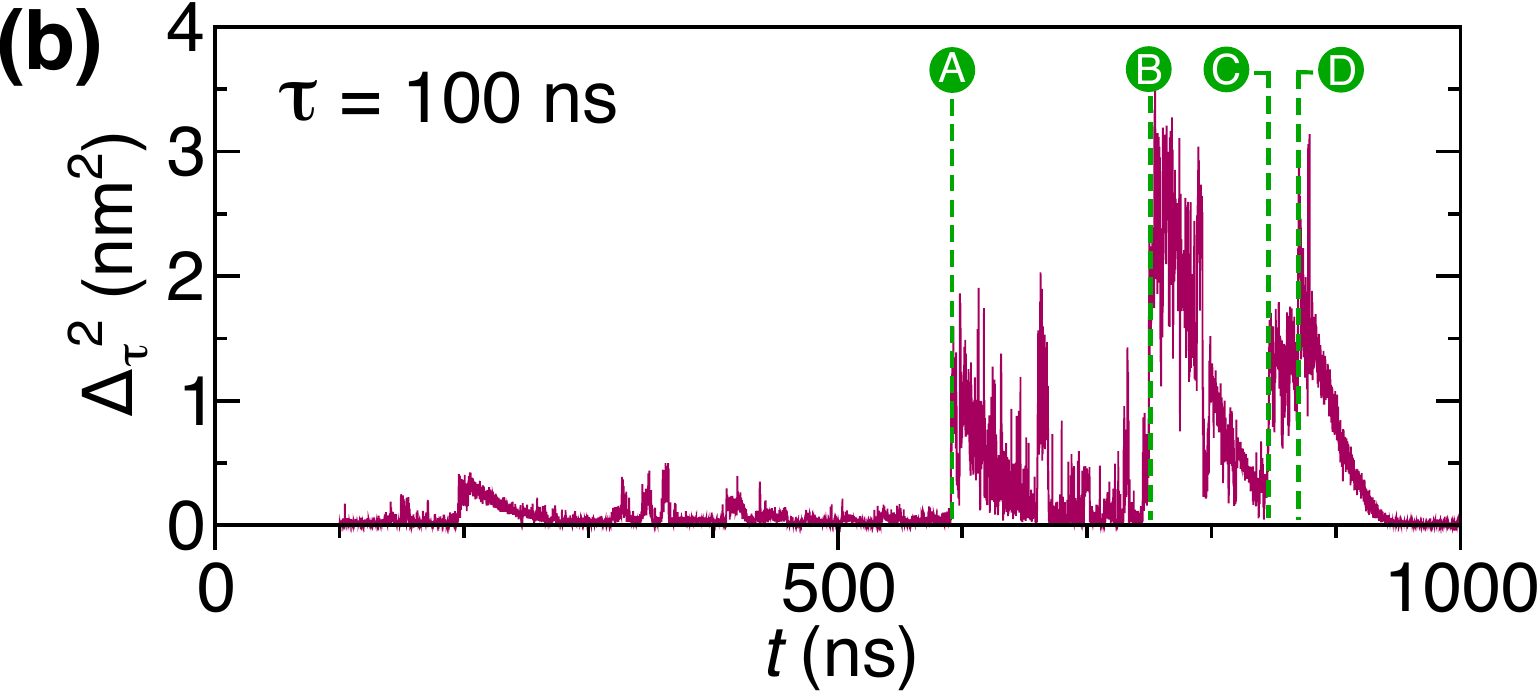}
\end{center}\end{minipage}
\caption{(a) A trajectory sequence of the penetrant NP$^0$ projected on the $xy$ plane (color coded from blue at $t=0$ to red at $t=1,000$~ns). Identified hopping transitions [see panel~(b)] are marked by letters.
(b)~The square displacement $\Delta_{\tau}^2(t)$ with $\tau=100$~ns (cf.\ \Eq~\ref{eq:DeltaTau}) of the above trajectory sequence. The identified hopping transitions (see Methods) are denoted by green vertical lines and marked also in the trajectory in (a).}
\label{fig:traj}
\end{center}\end{figure}

An example of a diffusion trajectory is shown in \Fig~\ref{fig:traj}a (projected on the $xy$ plane)  for the case of NP$^0$. The trajectory exhibits individual localized states, where the particle dwells for a sufficient amount of time without considerable migration. After some time the particle suddenly performs a larger jump to a different location where it gets `localized' into another dwelling state.
This is a characteristics of hopping diffusion~\cite{takeuchi1990jump, muller1991diffusion}, which has been identified also experimentally in collapsed PNIPAM-based gels~\cite{ghugare2010structure,philipp2014molecular} and 
  is observed quite generally in simulations of amorphous melts and glassy polymer matrices~\cite{muller1991diffusion,muller1992computational,gusev1994dynamics, muller1998molecular, fritz1997molecular, hahn1999new, hofmann2000molecular, kucukpinar2003molecular, mozaffari2010molecular}.
  

To quantitatively characterize the hopping events, we define the displacement of the particle from the preceding mean position as
\begin{equation}
\Delta_{\tau}^2(t)=\left[\overline{\Av r}({t-\tau, t})-\Av r(t)\right]^2
\label{eq:DeltaTau}
\end{equation}
The first term in the brackets is the time-averaged position of the particle during the time window $(t-\tau,t)$, whereas the second term is the position of the particle at time $t$.
The quantity $\Delta_{\tau}^2$ thus measures the square displacement of the particle from its mean position during the preceding time $\tau$. 
We plot the displacement $\Delta_\tau^2(t)$ with $\tau=$~100~ns for the NP$^0$ trajectory [presented in panel (a)] in \Fig~\ref{fig:traj}b.
During the dwelling states, $\Delta_\tau^2$ fluctuates around small values, reflecting the degree of the particle's fluctuations within a cavity. As the particle hops from one cavity to another, $\Delta_{\tau}^2$ exhibits an abrupt jump, followed by a decay on the time scale of $\tau$. Note that a hopping transition is much shorter than a typical residence time in the cavities. 
In a case the particle performs only a larger displacement fluctuation and promptly jumps back into the previous dwelling location, which is not considered as a hopping transition, $\Delta_{\tau}^2$ exhibits a $\delta$-like spike. This property enables us to distinguish hopping events from temporal displacement fluctuations. By using simple numerical criteria (see Methods), we isolate the hopping transitions from the trajectory, which are indicated by green vertical lines and annotated by letters A--D in \Fig~\ref{fig:traj}b, and correspondingly denoted also in panel~(a).

\begin{figure}[h]\begin{center}
\begin{minipage}[b]{0.48\textwidth}\begin{center}
\includegraphics[width=\textwidth]{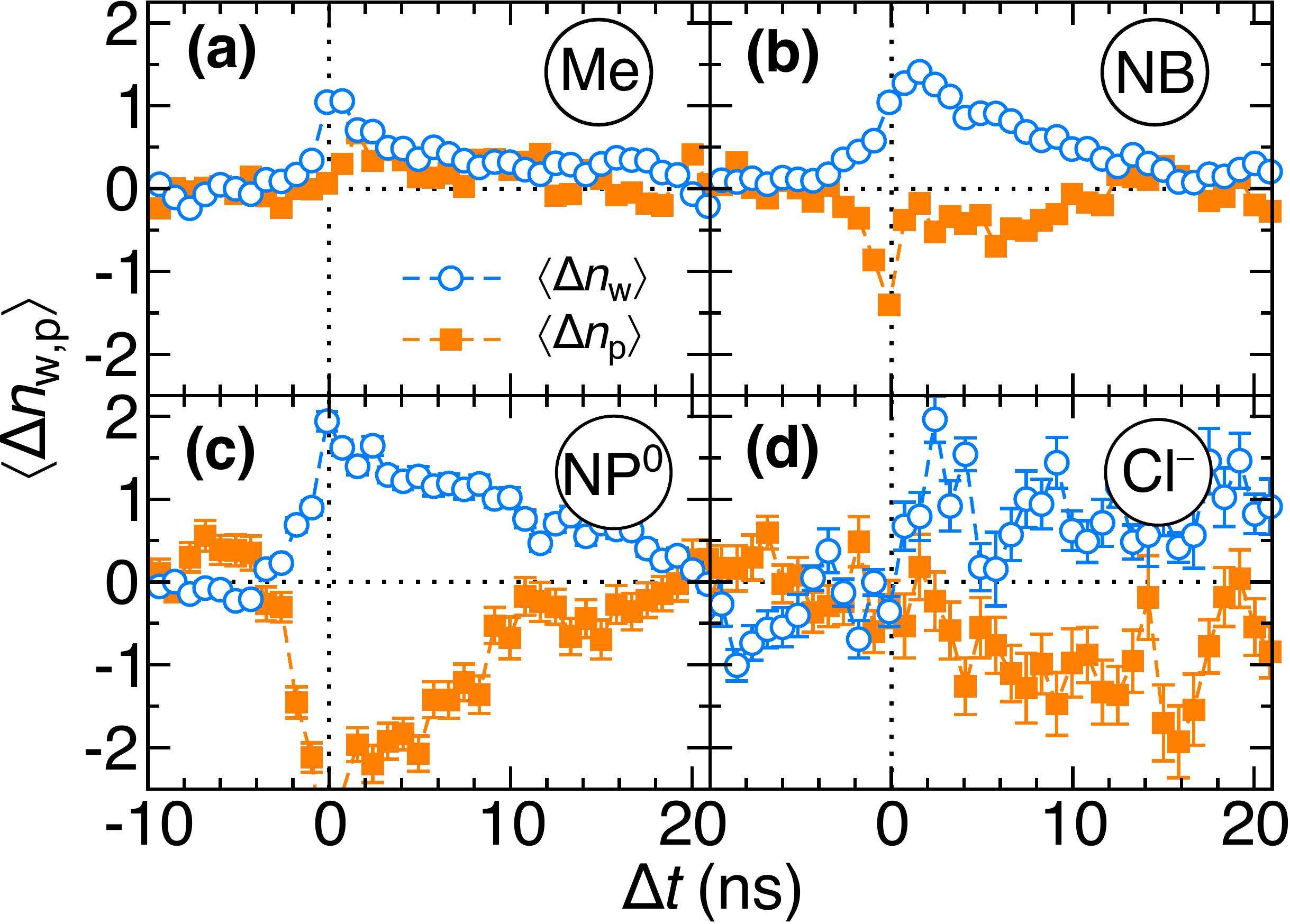}
\end{center}\end{minipage}
\caption{Mean changes in the hydration $n_\trm w$ (blue circles) and polymer coordination number $n_\trm p$ (orange squares) of diffusing particles during the hopping transition states.  The midway of the transition state, defined at $\Delta t=0$, is depicted by a dotted vertical line. The initial (reference) state set at $\Delta t = -10$~ns is conveniently defined to be $\Delta n_{\rm w,p}=0$.}
\label{fig:transition}
\end{center}\end{figure}

Having identified and characterized the hopping transitions, we are now able to track the changes in the hydration, $n_\trm w$, and the polymer coordination, $n_\trm p$, numbers during the transitions. Here, $n_\trm p$ counts, analogously as $n_\trm w$, the number of the polymer heavy atoms within  the cut-off $r_\trm{c}=0.54$~nm.
\Figure~\ref{fig:transition} shows the resulting averaged time evolution of the changes in
both numbers during the hopping events.
As representative examples we show cases of (a)~methane (small hydrophobe), (b)~nitrobenzene (large hydrophobe), (c)~nitrophenol (hydrophilic solute), and chloride (charged solute) in (d).
Universally, in all the cases, the penetrants get more hydrated during a hopping transition (observed jump in $n_\trm w$), which is for larger penetrants accompanied by a `detachment' from the polymer (drop in $n_\trm p$). Both quantities relax back to their respective mean values thereafter on a time scale of around 10~ns. 
\chgA{These main features are not qualitatively affected by the particular choice of the cut off (see \SItext).}
The results convey a clear three-phase pattern of the hopping mechanism: (i)~by thermal fluctuations, a transient channel opens, creating a water pathway from one cavity to an adjacent one, (ii)~the penetrant hops through the channel, (iii)~the channel closes.
This pattern, which appears to be quite general, has been found in other amorphous polymers as well~\cite{takeuchi1990jump, sok1992molecular,muller1993gas,muller1993cooperative,gusev1993dynamics,gusev1994dynamics, fritz1997molecular, muller1998diffusion}. 
The results imply that particles in our system universally perform hops not through dry regions of the polymer phase (even if they are non-polar), but through transient water channels instead.


\begin{figure}[h]\begin{center}
\begin{minipage}[b]{0.48\textwidth}\begin{center}
\includegraphics[width=\textwidth]{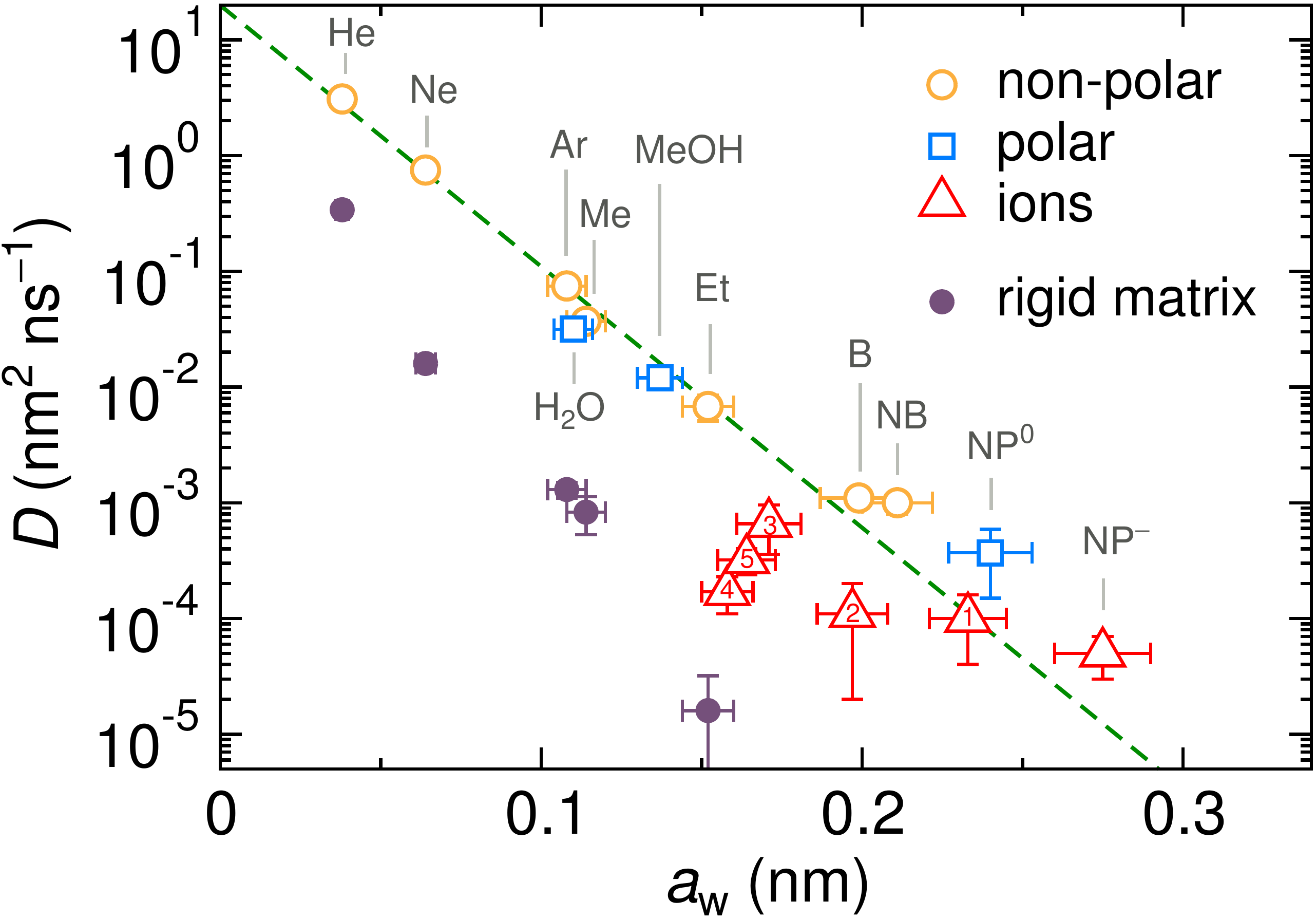}
\end{center}\end{minipage}
\caption{Long-time self-diffusion coefficients of penetrant molecules in the collapsed PNIPAM polymer at 340~K versus their Stokes radii in water. The fit of \Eq~\ref{eq:D_empirical} to the neutral penetrants (non-polar and polar) is shown by a green dashed line. Ions (red triangles, 1\,=\,Na$^+$~\cite{jorgensenFFNaCl}, 2\,=\,Na$^+$~\cite{aqvist1990ion}, 3\,=\,Cl$^-$~\cite{jorgensenFFNaCl}, 4\,=\,I$^-$~\cite{jorgensenFFI}, 5\,=\,I$^-$~\cite{dang1992nonadditive, rajamani2004size}) are not described well by the fit. The filled purple symbols correspond to the diffusion coefficients of non-polar penetrants in the case of completely immobilized PNIPAM chains.}
\label{fig:D}
\end{center}\end{figure}

We thus arrive at a central question of this study: How does the long-time diffusion coefficient depend on the particle size and type?
As is often the practice in the literature~\cite{amsden1998solute}, we express the sizes of the penetrants in terms of their hydrodynamic Stokes radii in pure water, $a_\trm w$ (see Methods). We plot the long-time self-diffusion coefficients in the collapsed polymer at 340~K versus the size of the penetrants  in \Fig~\ref{fig:D}. The results display a dramatic, five orders of magnitude large decrease in the diffusion coefficients as the penetrant size increases by a factor of~7. 
Orders of magnitude decrease of diffusion coefficients of polar molecular penetrants in a collapsed PNIPAM has been observed also in experiments~\cite{zhang2002diffusion}.
The results show that the diffusion coefficients of the non-polar and polar neutral molecules roughly follow the same trend, which can be described by an exponential function
\begin{equation}
D=D_0 \,\rme^{-a_\trm{w}/\lambda}
\label{eq:D_empirical}
\end{equation}
The best fit of this equation to the MD results for the neutral penetrants yields the decay length $\lambda=$~0.019~nm.
Based on these results, the behavior of the polar molecules does not differ significantly from the non-polar ones, indicating that the role of polarity in the diffusion plays a minor role. On the other hand, the diffusion of ions (shown by red triangles) clearly deviates from the trend of the neutral molecules of similar size. Notably, in the latter case, strong Coulomb interaction and the hydration seem to add an important contribution to the diffusion.

The significance of the fluctuating polymer matrix can be furthermore demonstrated by additional simulations of the same system but with rigid (\ie, completely immobilized) polymer chains~\cite{sok1992molecular,sentjabrskaja2016anomalous}. The resulting diffusivities of selected penetrants are shown by filled purple symbols in \Fig~\ref{fig:D}. In the rigid case, the diffusion plummets by an order of magnitude for the smallest penetrant helium, and even more for larger penetrants! Hence, the Boltzmann probability of crossing a barrier in the rigid matrix is far lower than in the case of mobile chains, which is additionally confirming that particle jumps are exceedingly facilitated by an elastic response of the matrix, which enables the openings of channels.

%

\subsection*{Activated hopping: Analysis of temperature and hydration effects}
%

\begin{figure}[h!]\begin{center}
\begin{minipage}[b]{0.41\textwidth}\begin{center}
\includegraphics[width=\textwidth]{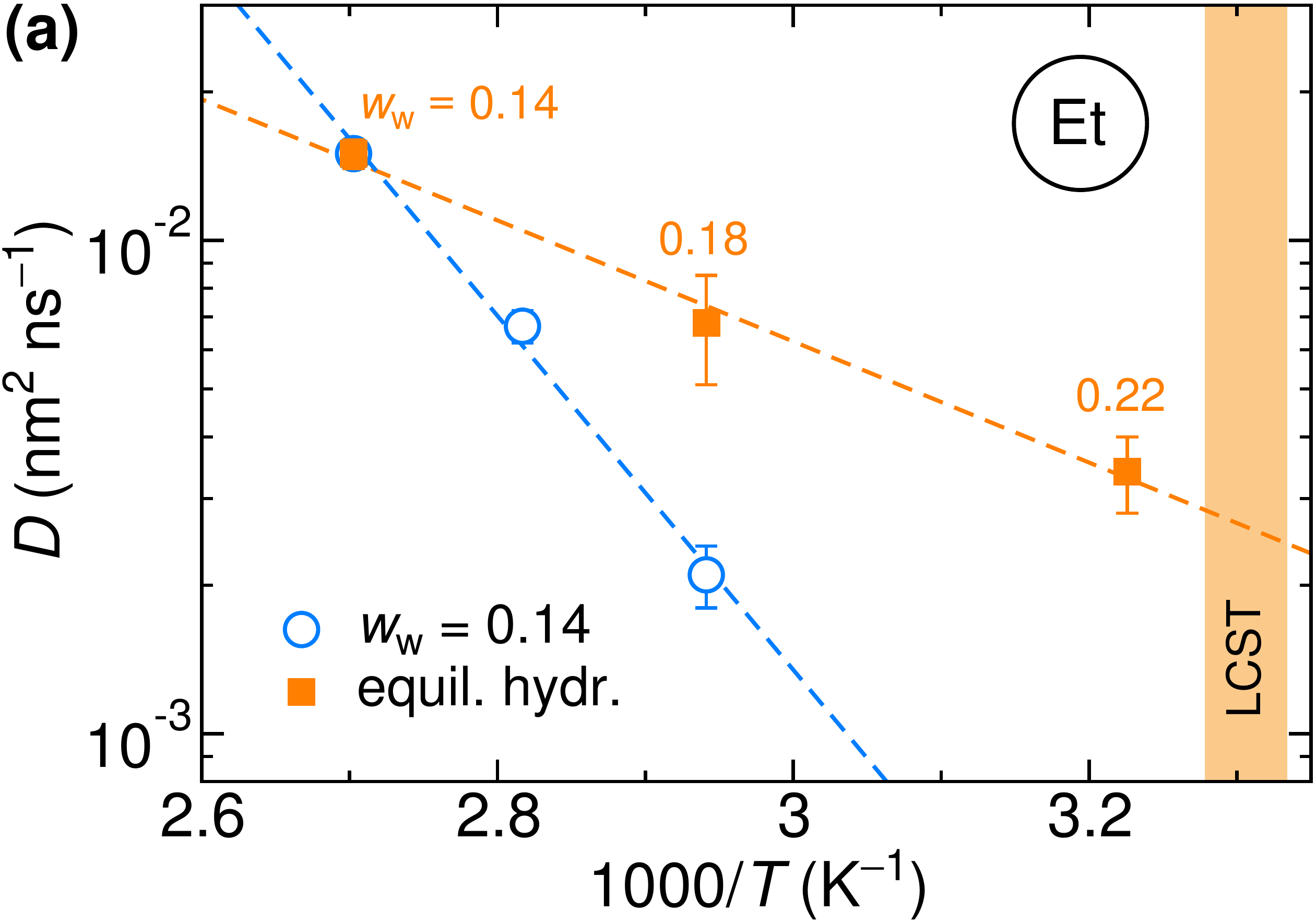}
\end{center}\end{minipage}\vspace{1ex}
\begin{minipage}[b]{0.43\textwidth}\begin{center}
\includegraphics[width=\textwidth]{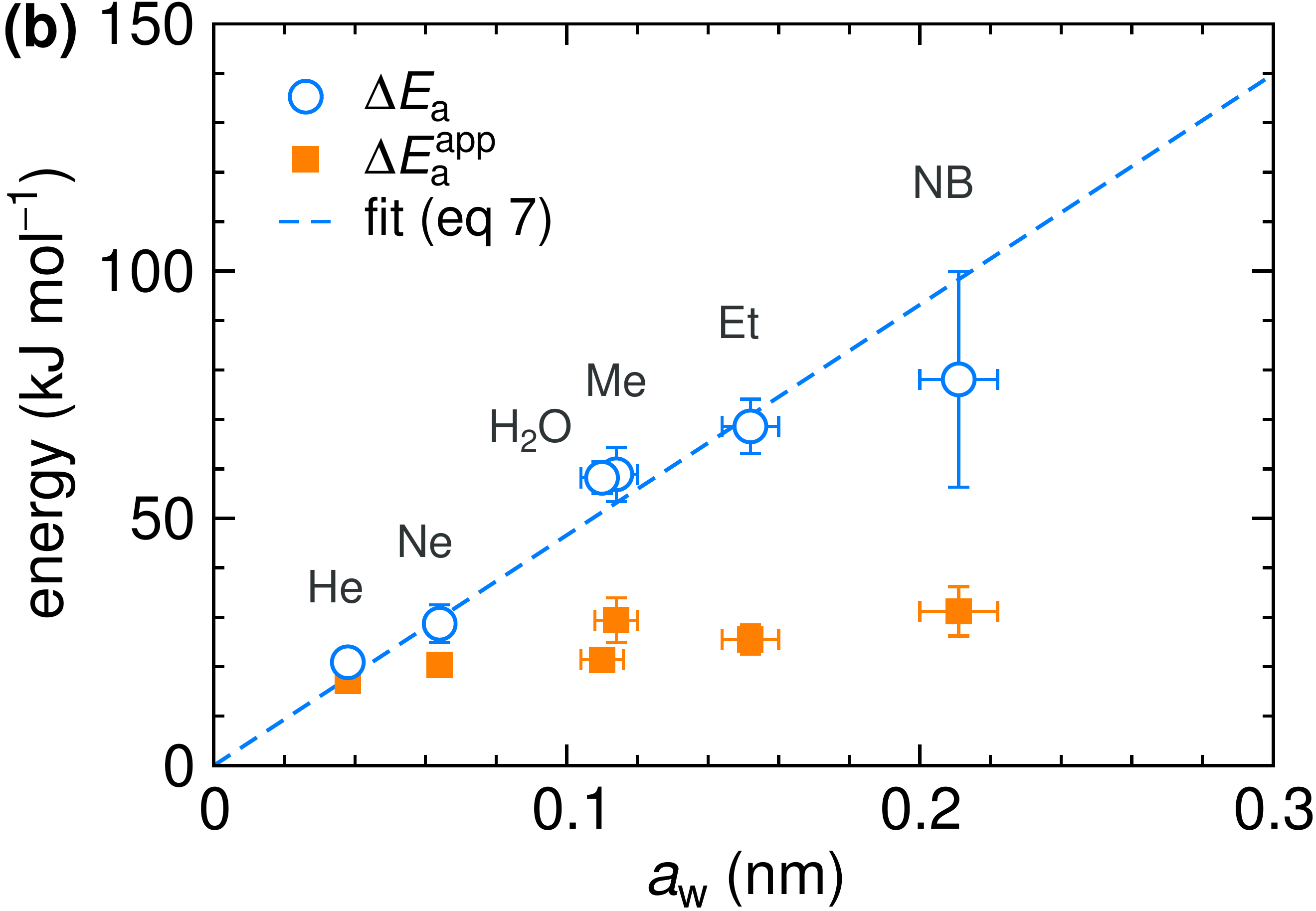}
\end{center}\end{minipage}
\caption{
(a)~Arrhenius plot of the diffusion coefficient for ethane. The filled square symbols represent the results in the polymers with equilibrated hydration at each temperature (water fractions are indicated at each symbol). The open circles correspond to the system with a fixed hydration level of $w_\trm w=$~0.14. 
The dashed lines are the fits of the Arrhenius equation (\Eq~\ref{eq:Arrhenius}) to the MD data points.
(b)~Energy barriers $\Delta E_\trm a$ and $\Delta E_\trm a^\trm {app}$ for the selected penetrants evaluated from the fits by using \Eqs~\ref{eq:dEw} and \ref{eq:dEmu}, respectively. The blue dashed line is the fit of \Eq~\ref{eq:dEwaw} to the data points for $\Delta E_\trm a$.
}
\label{fig:D2}
\end{center}\end{figure}

We now address  the influences of temperature and hydration on the diffusion. Both effects are demonstrated on the case of ethane by two scenarios in \Fig~\ref{fig:D2}a in a form of an Arrhenius plot.
In the first scenario (blue circles), the water content in the polymer is fixed at $w_\trm w =$~0.14 at all temperatures. Here, the expected diffusivity increment with temperature can be entirely ascribed to thermal effects.
The second scenario (orange squares) corresponds to the polymer phase with the water component in chemical equilibrium with bulk water at each temperature, where the water fraction $w_\trm w$ equals the values shown in \Fig~\ref{fig:rwT}c. 
 Also in this case, the diffusion coefficient rises with temperature, albeit significantly less than in the first scenario. Namely, the thermal effects in this case mix with the effects of temperature-dependent (de)hydration.
Hence, at given temperature, a more hydrated polymer provides higher diffusivity than a less hydrated one.

The temperature dependence of activated diffusion can be analyzed by the Arrhenius relation
\begin{equation}
D=\tilde D\,\rme ^{-\Delta E_\trm{a}/\kB T}
\label{eq:Arrhenius}
\end{equation}
Assuming that the prefactor is independent of temperature, the energy barrier then follows as
\begin{equation}
\Delta E_\trm{a}=\kB T^2\left(\frac{\partial\, \ln D}{\partial T}\right)_{w_\trm{w}}
\label{eq:dEw}
\end{equation}
where the partial derivative is taken at constant water fraction $w_\trm w$.
  \Figure~\ref{fig:D2}b shows evaluated energy barriers $\Delta E_\trm a$ for selected uncharged penetrants versus their sizes. Clearly, the results suggest a linear dependence
 \begin{equation}
\Delta E_\trm{a}=C_E a_\trm w
\label{eq:dEwaw}
\end{equation}
plotted by a dashed line with the fitting parameter $C_E=470(30)$~kJ\,mol$^{-1}$nm$^{-1}$.
 
In the other scenario, where the water component is in equilibrium with bulk reservoir, the resulting temperature dependence of the diffusivity is more gradual (\Fig~\ref{fig:D2}a), with the slope
\begin{equation}
\Delta E_\trm{a}^\trm{app}=\kB T^2\left(\frac{\partial\, \ln D}{\partial T}\right)_{\mu_\trm{w}}
\label{eq:dEmu}
\end{equation}
This quantity represents an {\it apparent} energy barrier that lumps together the contributions of the thermal influence and the variable hydration into a single macroscopic parameter. The values of $\Delta E_\trm{a}^\trm{app}$, shown in \Fig~\ref{fig:D2}b, are much less sensitive to the particle size than $\Delta E_\trm{a}$. Consequently, an important finding is that the variable hydration contributes a strongly compensating component to the apparent energy barrier for diffusion.

\subsection*{Activated hopping: Discussion}
The rate-determining step in the hopping diffusion is the opening of a channel, which is associated with a free energy barrier $\Delta F_\trm{a}$ and can be via Boltzmann probability related to the diffusion coefficient as $D\sim \exp(-\Delta F_\trm{a}/\kB T)$. In conjunction with the empirically obtained diffusion relation (\Eq~\ref{eq:D_empirical}), this implies
\begin{equation}
\Delta F_\trm{a}(a_\trm w)=\frac{\kB T}{\lambda} a_\trm w
\label{eq:dF}
\end{equation}
This linear dependence of the free energy barrier on the particle size represents \chgA{a special case of possible scenarios predicted by various theories. Most of the now classical theories that are based on activated diffusion describe regimes with either square~\cite{peppas1983solute, amsden1999obstruction, amsden2002modeling} or cubic~\cite{vrentas1977diffusion, vrentas1985free, fujita1991notes, dell2013theory} scaling. However, these theories have been developed assuming a water-swollen network and bigger penetrants, such as enzymatic drugs or nanoparticles.} A possible linear dependence of the free energy barrier has recently been theoretically envisioned in scaling theories for particle mobility in dense polymer solutions~\cite{cai2011mobility, cai2015hopping} and in dense liquids by using a self-consistent cooperative hopping theory~\cite{zhang2017correlated}.
According to these studies, the actual linear scaling regime seems to depend in a very complex way on various system parameters~\cite{cai2015hopping} that are still under debate~\cite{zhang2017molecular, zhang2018coarse}. Note also that these statistical mechanics theories based on ideal-chain models of polymers and no explicit hydration effects predict purely entropic free energy barriers for hopping, in stark contrast to the strong $T$-dependence of diffusion observed in our simulations.

Previous atomistic simulation studies~\cite{kucukpinar2003molecular, mozaffari2010molecular} support a diffusion relation that decays exponentially with the {\it square} of the effective penetrant size, $\ln\, D \sim -a_\trm{eff}^2$, however, one should keep in mind the significance of a fit to given numerical data. If a span of particle sizes is not large enough, several functional forms can be fitted to the same data points within a given accuracy.
In our results in \Fig~\ref{fig:D} the ratio between the smallest and the largest penetrant is around 7, which is much more than in older atomistic studies, thus providing higher significance to the relation given by \Eq~\ref{eq:D_empirical}. 


In addition, our study points to the crucial role of the water component in the polymer.
We found that, at given temperature, a more hydrated polymer provides higher diffusivity than a less hydrated one.
   A similar effect is known in some hydrophilic synthetic polymers (\eg, polyvinyl alcohol, used in food industry for packaging materials) that possess high barrier resistance for diffusion of various gases under dry conditions, but become drastically more permeable in humid environments~\cite{hernandez1994effect, zhang2001permeation, karlsson2004molecular}.
The phenomenon is commonly ascribed to the sorbed water, which acts as a plasticizer and reduces the tensile strength of the polymer by softening interactions between the chains~\cite{muller1998diffusion, tamai1998permeation, karlsson2004molecular}.
This means that less hydrated polymer systems have higher free energy barriers for channel formations, and according to \Eq~\ref{eq:dF} this implies that they should exhibit a smaller diffusion decay length $\lambda$. Indeed, as we show in \SItext, the diffusion coefficients of selected penetrants in a less hydrated polymer decay faster with their size. This can be viewed within a simplified picture as follows: Firstly, in more hydrated polymers, the average separation between the chains is larger (seen as a decreased PNIPAM partial density in \Fig~\ref{fig:rwT}b), which have therefore more freedom for fluctuations and thereby creating channel openings more readily. Secondly, in more hydrated cases with larger and more abundant water clusters, the separations between neighboring clusters are smaller on average, consequently the activated transient channels can be shorter.

Finally, our analysis conveyed another important message: The energy barrier $\Delta E_\trm a$ is proportional to the particle size. This has an interesting consequence that larger diffusing molecules are more affected by temperature changes at fixed hydration of the polymer than smaller ones. 
This effect can be used as an additional means in tailoring and tuning the selectivity of a molecular transport through hydrogels by external stimuli.


\section*{Conclusions}
We have presented a detailed all-atom simulation analysis of penetrant diffusion in a collapsed PNIPAM polymer phase in water. Water has been found to structure in fractal-like clusters sorbed between the voids made by the polymeric chains. The diffusion advances via the hopping mechanism, in which a penetrant resides for longer time in a local cavity and suddenly performs a longer jump into a neighboring cavity through a transient water channel that forms between the chains. 

We found that the diffusion heavily depends on the temperature and the hydration level of the polymer. These two effects are typically coupled, since the thermo-responsive nature of PNIPAM directly impacts the affinity to water, thereby regulating its hydration through an outer water reservoir. By systematic and careful simulation approaches, we were able to separate both effects and demonstrate plasticizing effects of water on the diffusion. 

Furthermore, the penetrant molecules in our study extend almost over an order of magnitude in their sizes and almost over five orders of magnitude in the resulting diffusivities. This allowed us to  formulate a reliable and statistically significant relation between the diffusion coefficients and the sizes of  the penetrants. We find that the diffusion of non-ionic penetrants follow a universal, exponential dependence on the size. The diffusion of ions, on the other hand, deviates from this relation, probably due to additional Coulomb interactions with the polymer chains.  

The outcome of this work seriously challenges theoretical understanding of diffusion in such systems. As we have seen in our simulations, the dynamics of small and mid-sized penetrants is coupled to the structural relaxation of the polymer, which has far-reaching consequences and it complicates theoretical modeling. That means, it is not sufficient to treat a collapsed polymer with rigid obstruction models, which are very popular due to their simplicity~\cite{amsden1998solute, masaro1999physical}. Moreover, the hopping cannot be explained by hydrodynamic forces.
In addition, the diffusion in our system is thermally activated, which manifests in large energy barriers that scale linearly with the size of the penetrant. As we have demonstrated, the hydration effects stemming from the clustered water, together with the thermo-responsive nature of the polymer, are crucial elements that have to be included in future theoretical considerations.

We believe that our findings will stimulate new theoretical efforts into this problem.
 Understanding the transport mechanisms inside not only PNIPAM but also other responsive hydrogels is important for the rational design of novel materials. 
\chgA{We are currently performing a follow-up study of solvation properties of penetrants in these systems, which will complement our understanding of penetrant interactions with collapsed hydrogels.}

%



\small
\section*{Methods}
\subsection*{Atomistic model}
We utilize an atomistic model of PNIPAM polymer in the presence of explicit water.
The PNIPAM chains are composed of 20 monomeric units with atactic stereochemisty (\ie, with random distribution of monomeric enantiomers along the chain). 
To describe water in the simulations we use the SPC/E water model~\cite{spce}. 

For PNIPAM polymers we first tested the standard OPLS-AA~\cite{opls1988} force field,
which is among the most popular ones for PNIPAM simulations~\cite{walter-PNIPAM2010, vegtJPCB2011, stevens_Macro2012, mukherji2013coil,  chiessi2016influence, rodriguez2014direct, adroher2017conformation}.
Since this force field did not yield satisfactory hydration results in the collapsed state (see \Fig~\ref{fig:rwT}c) and also due to revealed issues with its thermo-responsive properties in the recent literature~\cite{kang2016collapse, botan2016direct}, we use its recent modification 
with recalculated partial charges by Palivec et al.~\cite{palivecheyda2018}.
As has been demonstrated, the novel forcefield exhibits thermo-responsive properties of a single PNIPAM polymer much closer to experimental observations.
\chgA{For the neutral penetrant molecules, we use the OPLS-AA force field~\cite{opls1988, priceOPLS2001}, which keeps our model on the generic level and which sufficiently captures the hydration properties in combination with the SPC/E water model~\cite{hess2006hydration}.} For the deprotonated ion NP$^-$, which is not provided within OPLS-AA, we used the partial charge parameterization `OPLS/QM1' from Ref.~\citen{kanduc2017selective}.
For the monatomic ions we employ the parameters from the Jorgensen force fields~\cite{jorgensenFFNaCl, jorgensenFFI}. For comparison, we additionally use the \AA qvist \cite{aqvist1990ion} force field for Na$^+$ and Dang et al.~\cite{dang1992nonadditive, rajamani2004size} for I$^-$.

\subsection*{Simulation procedures}
In the first part of the study, we assembled 48 polymer chains in a slab-like structure that extended across the $x$ and $y$ box dimensions with a finite thickness (approx.\ 3--4~nm) in $z$-direction.
Initially, the polymers were only loosely arranged in a slab-like assembly and solvated with water. The equilibration steered the assembly into a more compact structure, thereby making two well separated phases of water-only (supernatant) and the polymer-rich (precipitant) slab along $z$-axis as shown in the snapshot in \Fig~\ref{fig:rwT}a.
For equilibration purposes, we performed simulated annealing where the temperature was linearly decreased by the thermostat from 450~K down to a target temperature (\ie, 310~K, 340~K, or 370~K) on a time interval of 100~ns. After that, the equilibration continued at the target temperature until the water amount in the gel phase, which we monitored, equilibrated (see \SItext).
The necessary simulation times for equilibration depend significantly on temperature and are around 2,000~ns, 1,000~ns, and 500~ns for $T$\,$=$\,310~K, 340~K, and 370~K, respectively.
Notably, much longer equilibration times are needed at lower temperatures due to slower kinetics.
In order to improve sampling statistics, we averaged the results over four independent sets of simulations for 310~K and 340~K, and over two sets for 370~K.

In the second part we set up single PNIPAM-phase simulations. Here, 48 atactic PNIPAM polymers in a cubic box were mixed with a certain number of water molecules. 
We equilibrated the systems via the annealing simulations with linearly decreasing temperatures from 900~K down to target temperatures on a time interval of 100~ns. After that, the equilibration continued for another 50--100~ns. 
The resulting equilibrated structures (with an example shown in \Fig~\ref{fig:gelly}) were then used as initial configurations for all further analyses in this work.

To study diffusion, we inserted 10--15 penetrant molecules of the same kind at random positions into the equilibrated polymer structures (using 2--4 independent replicas).
When simulating ions, we estimate the Bjerrum length to 5--6~nm (assuming the dielectric permittivity of a collapsed PNIPAM in water to be around 10~\cite{sasaki1999dielectric}), which could lead to mutual influences between the ions due to strong Coulomb interactions. To reduce the effects, we restricted the number of ions in a simulation box to three, which on the other hand compromised the statistics. The net charge was compensated by applying a uniform neutralizing background charge.
The simulation production runs spanned from 300~ns for the smallest penetrants (He, Ne), to around 1,000~ns for medium-sized (methane, ethane), and up to \chgA{8,000~ns} for the largest ones (NB, NP$^0$) and ions. \chgA{Such long simulation times for the latter penetrants were required in order to observe at least several hopping events.}


\subsection*{Simulation details}
The molecular dynamics~(MD) simulations were performed with the GROMACS~5.1 simulation suite~\cite{gromacs,gromacs2013} in the constant-pressure (NPT) ensemble, where the box sizes are independently adjusted in order to maintain the external pressure of 1~bar via Berendsen barostat~\cite{berendsenT} with the time constant of 1~ps. 
The system temperature was maintained by the velocity-rescaling thermostat~\cite{v-rescale} with a time constant of 0.1\,ps.
The Lennard-Jones~(LJ) interactions were truncated at 1.0~nm. Electrostatics was treated using Particle-Mesh-Ewald (PME) methods~\cite{PME1,PME2} with a 1.0~nm real-space cutoff.

\subsection*{Determination of hopping transitions}

Identifying the hopping transitions from $\Delta_{\tau}^2(t)$ is related to the step detection problem in a noisy signal and is thus a matter of subjective criteria.
Due to the noisy nature, we scanned through $\Delta_{\tau}^2(t)$ by defining two moving average values, a lagging average $\Delta_{\tau,-}^2(t)\equiv\langle\Delta_{\tau}^2\rangle_{t-t_1,t}$, which is the mean value of $\Delta_{\tau}^2(t)$ in the time  window $(t-t_1,t)$, and similarly an advancing average $\Delta_{\tau,+}^2(t)\equiv\langle\Delta_{\tau}^2\rangle_{t,t+t_1}$ in the window $(t, t+t_1)$. Obviously, a step in $\Delta_{\tau}^2$ occurs when the ratio $r(t)\equiv\Delta_{\tau,+}(t)/\Delta_{\tau,-}(t)$ exhibits a local peak. We specified two criteria that have to be fulfilled in order to consider an identified peak as a hopping event.
First, the local peak in $r(t)$ should correspond to the maximal value in the time window $(t-t_1, t+t_1)$. This assumes that at most one hopping event can occur in the given time window of length $2t_1$. 
Second, the jump should fulfill the condition $\Delta_{\tau,+}^2(t)-\Delta_{\tau,-}^2(t)> {\delta_\trm{min}}^2$, that is, the particle has to jump by more than a threshold distance $\delta_\trm{min}$. 
For all the analyzed penetrants in \Fig~\ref{fig:transition}, we chose averaging interval length $t_1=$~5~ns (which is considerably longer than temporal displacement fluctuations and much shorter than a typical residence time of a penetrant in a cavity) and the threshold $\delta_\trm{min}=$~0.75~nm (which roughly corresponds to an estimated inter-chain distance, thus to a typical cavity size).



\subsection*{Long-time diffusion}
A conventional way to analyze the particle dynamics in computer simulations, is to evaluate mean square displacement~(MSD) of the particles,
\begin{equation}
\Delta^2(\tau)=\left\langle |\Av r_i(t+\tau)-\Av r_i(t)|^2\right\rangle_{t,i}
\end{equation}
where $\Av r_i(t)$ is the position of the particle $i$ at time $t$. When evaluating an MSD, 
the center-of-mass motion of the whole system should be removed from the displacement vector $\Av r_i(t+\tau)-\Av r_i(t)$.
However, the diffusion of particles in crowded environments often behaves anomalously at short time scales, as it is coupled to the segmental dynamics of polymers, and MSD follows a more general power-law pattern $\Delta^2(\tau)\sim \tau^\alpha$, where the scaling exponent $\alpha$ can deviate from unity~\cite{bouchaud1990anomalous, muller1992computational, metzler2000random, sokolov2012models, ernst2014probing, metzler2014anomalous, ghosh2015non}. Only at sufficiently long observation times the MSD behavior crosses over to normal (Brownian) diffusion with $\alpha=1$. Since we are interested solely in the normal (long-time) diffusion, it is important to assess the crossover time for each individual MSD.
As we show in \SItext, the crossover time depends on the size of the particle, spanning from 
 around 1~ns for the smallest one (helium) and up to several 100~ns for larger molecules. This analysis indicates that for the largest considered penetrant particles, trajectory lengths of several \si\micro s are needed in order to properly evaluate the diffusion coefficients. 
 As noted before~\cite{gusev1994dynamics}, this stringent verification of the crossover time was often lacking in many previous evaluations of MD results.
Once the crossover time has been determined, the diffusion coefficient can be calculated from   the Einstein relation, viz.
\begin{equation}
D=\lim_{\tau\to\infty}\frac 16 \frac{\rmd}{\rmd \tau}\Delta^2(\tau)
\label{eq:limD}
\end{equation}
which we achieve by a linear fit of the MSD in the long-time limit.

In our simulations, we also evaluated the MSD of individual polymer chains. On the time scale of the simulations, the diffusion of the polymer chains is negligible compared to the diffusion of the penetrants, which indicates that the penetrant diffusion is not related to the diffusion of the whole network. 

\subsection*{Stokes radii in water}
The dominant factor that governs the diffusion is the size of the particle~\cite{gusev1994dynamics}, whereas other factors, such as the shape and polarity, are typically of secondary importance.
As is often the practice in the literature~\cite{amsden1998solute}, we express the sizes of the penetrants by their Stokes hydrodynamic radii in pure water, $a_\trm w$, defined via the Stokes--Einstein equation, 
\begin{equation}
D_\trm w=\frac{\kB T}{6\pi \eta a_\trm w}
\label{eq:SE}
\end{equation}
where $D_\trm w$ is the diffusion coefficient in water and $\eta$ the water viscosity.
The Stokes radius represents a suitable measure of the effective particle size, as it captures the `bare' particle size together with its hydration shell. The latter one is relevant in our case, since the penetrants, as we have seen, diffuse predominantly through transient water channels. In addition, the Stokes radius is a well-defined quantity in experiments and simulations. Therefore, we used the Stokes--Einstein equation \ref{eq:SE} to determine Stokes hydrodynamic radii $a_\trm w$ of the molecules in pure water, which required the evaluation of the  $D_\trm w$ and $\eta$.
The latter was computed with the standard procedure of transverse current correlation function~\cite{palmer1994transverse, hess2002determining} from an independent simulation of pure water at 340~K. The obtained value $\eta=0.39(2)$~mPa$\cdot$s agrees very well with previously reported 0.38~mPa$\cdot$s for the SPC/E water at this temperature~\cite{markesteijn2012comparison}.
The diffusion coefficients were obtained from the MSDs~(\Eq~\ref{eq:limD}) of the molecules in a water box.
Since hydrodynamic interactions are long range ($\sim 1/r$), they lead to  effective coupling between the molecule, the solvent, and the periodic images.
The evaluated {\em apparent} diffusion coefficients $D_\trm{w}^\trm{app}(L)$ are therefore box-size dependent. The finite-size effects on the diffusivity can be estimated by the Ewald summation of the Oseen tensor over the periodic images, which enables the evaluation of the actual diffusion coefficient $D_\trm{w}$ as~\cite{yeh2004diffusion,yeh2004system}
 \begin{equation}
D_\trm{w}=D_\trm{w}^\trm{app}(L)+\frac{\xi_\trm{EW}\kB T}{6\pi\eta L}
\label{eq:Dcorr}
\end{equation}
Here, $L$ stands for the box length and the constant $\xi_\trm{EW}=2.837$ stems from the summation over all periodic images.

\Equation~\ref{eq:Dcorr} is based on the first-order correction and therefore applicable only for $L\gg a_\trm{w}$~\cite{hasimoto1959periodic, yeh2004system}. In order to determine what is the minimal box size $L$ that enables the application of the correction given by \Eq~\ref{eq:Dcorr}, we performed a size-scaling analysis of $D_\trm{w}^\trm{app}(L)$ for water, NB, and I$^-$ (see \SItext). Based on the outcomes, we used box sizes of $L=6$~nm for the larger (aromatic) molecules and $L=4$--$5$~nm for smaller molecules.

The results for Stokes radii do not significantly depend on temperature in the range 310--370~K  (see ~\SItext). Some differences are observed only for He and Ne. 
However, we use the values of $a_\trm w$ obtained at 340~K for all the particles and regard them as fixed molecular parameters. 

\subsection*{Conflict of Interest} The authors declare no competing
financial interest.

\subsection*{Supporting Information Available}
Equilibration; hydration shell of methane; finite-size correction of diffusion in water; Stokes radii at different temperatures; short-diffusion diffusion analysis; diffusion at different temperatures and hydrations

\subsection*{Acknowledgments}
The authors thank Richard Chudoba, Vladimir Palivec, Jan Heyda, Sebastian Milster, Yan Lu, and Matthias Ballauff for useful discussions.
This project has received funding from the European Research Council (ERC) under the European Union's Horizon 2020 research and innovation programme (grant agreement n$^\circ$ 646659-NANOREACTOR).
The simulations were performed with resources provided by the North-German Supercomputing Alliance~(HLRN).

\setlength{\bibsep}{0pt}
\bibliography{literature}

\end{document}